\documentclass[twocolumn, superscriptaddress,nofootinbib,nofootinbib,preprintnumbers  ]{revtex4-1}   
\usepackage{graphicx}  
\usepackage{dcolumn}   
\usepackage{bm}        
\usepackage{amssymb, amsmath}
\usepackage[utf8]{inputenc}
\usepackage{newtxtext}
\usepackage{siunitx}

\def\beq{\begin{equation}}
\def\eeq{\end{equation}}
\usepackage[usenames,dvipsnames,svgnames]{xcolor}
\usepackage{hyperref}
\hypersetup{colorlinks, citecolor=black, linkcolor=black, urlcolor=black}

\usepackage{amsmath}
\usepackage{amsfonts}
\usepackage{amssymb}
\usepackage{bbm}                
\usepackage{graphicx, rotating}
\usepackage{epstopdf}
\usepackage{epsfig}
\usepackage{latexsym}
\usepackage{graphicx}
\usepackage{bbold}
\usepackage{color}
\usepackage{amsmath,amssymb}
\usepackage{lipsum}

\DeclareSIUnit\electronvolt{e\kern-.05em V}

\begin{document}

\preprint{MIT-CTP/5126}

\title{Reviving Millicharged Dark Matter for 21-cm Cosmology}

\author{Hongwan Liu}
\affiliation{Center for Theoretical Physics, Massachusetts Institute of Technology, Cambridge, MA 02139, U.S.A.}
\author{Nadav Joseph Outmezguine}
\affiliation{Raymond and Beverly Sackler School of Physics and Astronomy, Tel-Aviv University, Tel-Aviv 69978, Israel}
\author{Diego Redigolo}
\affiliation{Raymond and Beverly Sackler School of Physics and Astronomy, Tel-Aviv University, Tel-Aviv 69978, Israel}
\affiliation{Department of Particle Physics and Astrophysics, Weizmann Institute of Science, Rehovot 7610001,Israel}
\author{Tomer Volansky}
\affiliation{Raymond and Beverly Sackler School of Physics and Astronomy, Tel-Aviv University, Tel-Aviv 69978, Israel}

\begin{abstract}
The existence of millicharged dark matter (mDM) can leave a measurable imprint on 21-cm cosmology through mDM-baryon scattering. However, the minimal scenario is severely constrained by existing cosmological bounds on both the fraction of dark matter that can be millicharged and the mass of mDM particles. We point out that introducing a long-range force between a millicharged subcomponent of dark matter and the dominant cold dark matter (CDM) component leads to efficient cooling of baryons in the early universe, while also significantly extending the range of viable mDM masses. Such a scenario can explain the anomalous absorption signal in the sky-averaged 21-cm spectrum observed by EDGES, and leads to a number of testable predictions for the properties of the dark sector. The mDM mass can 
then lie between 10 MeV and a few hundreds of GeVs, and its scattering cross section with baryons lies within an unconstrained window of parameter space above direct detection limits and below current bounds from colliders. In this allowed region, mDM can make up as little as $10^{-8}$ of the total dark matter energy density. The CDM mass ranges from 10 MeV to a few GeVs, and has an interaction cross section with the Standard Model that is induced by a loop of mDM particles. This cross section is generically within reach of near-future low-threshold direct detection experiments.
\end{abstract}

\maketitle

\section{Introduction}

Dark Matter (DM)-baryon interactions can affect the thermal history after recombination~\cite{Tashiro:2014tsa,Munoz:2015bca}, possibly leading to observable deviations in the 21-cm global signal compared to the $\Lambda\text{CDM}$ prediction~\cite{Loeb:2003ya}.  Moreover, if the scattering cross section is enhanced at small relative velocities, its impact on the cosmological history is most prominent  between recombination and reionization, when the baryons are at their 
coldest. 
The most optimistic scenario for identifying new physics in 21-cm observables is therefore realized if the DM-baryon interactions are Rutherford-like, i.e.\ with a cross section scaling as  $v^{-4}_{\text{rel}}$, where $v_\text{rel}$ is the relative velocity between particles.  This may result from an exchange of a light mediator with  negligible mass compared to the typical exchange momentum in the scattering process. 

In most scenarios, however, the presence of a new light mediator is strongly constrained by a combination of fifth-force and stellar cooling bounds~\cite{Adelberger:2003zx,Hardy:2016kme,An:2014twa}.  Accounting for these constraints leaves little room for observable effects from DM-baryon scattering in 21-cm cosmology~\cite{Barkana:2018cct}, with the only exception being a DM that carries an effective electric charge $Q$, inducing DM interactions with the SM \footnote{In this paper, $Q$ is the charge measured in units of the electron charge so that the coupling of the DM to the SM photon reads $e Q A_\mu J_{\text{DM}}^{\mu}$. In the nearly massless dark photon case, one obtains $Q\sim\sqrt{\alpha_D \epsilon^2/\alpha_{\text{EM}}}$, with $\epsilon$ being the parameter controlling the mixing between the visible and dark sectors, while $\alpha_D=g_D^2/4\pi$ is defined by the minimal coupling to the DM current, $g_{D} A_\mu' J_{D}$. Alternatively, the DM electromagnetic charge can be assigned by coupling it directly to the SM photon.\label{eq:footnotedark}}.  
Even in this so-called \emph{millicharged dark matter} (mDM) scenario, which has been studied extensively in the literature~\cite{Tashiro:2014tsa,Munoz:2015bca,Munoz:2018pzp,Barkana2018,Barkana:2018cct,Berlin:2018sjs,DAmico:2018sxd,Liu:2018uzy,Mitridate:2018iag}, several stringent experimental constraints~\cite{Munoz:2018pzp,Barkana:2018cct,Berlin:2018sjs} apply.
Most notably, measurements of the CMB anisotropy power spectrum~\cite{Dubovsky:2003yn,Dolgov:2013una,Kovetz:2018zan,dePutter:2018xte,Boddy:2018wzy} and $N_\text{eff}$ constraints~\cite{Berlin:2018sjs,Creque-Sarbinowski:2019mcm} limit the mDM to a small region of parameter space, with mass $\SI{10}{\mega\eV} \lesssim m_\text{m} \lesssim \SI{40}{\mega\eV}$, charge $10^{-5} \lesssim Q \lesssim 10^{-4}$, and energy density between 0.01\% and 0.4\% of the total dark matter energy density. These severe restrictions make simple models of mDM unattractive as a significant source of cooling for baryons in the early universe.

\begin{figure*}[t]
\centering
\includegraphics[width=0.9\textwidth]{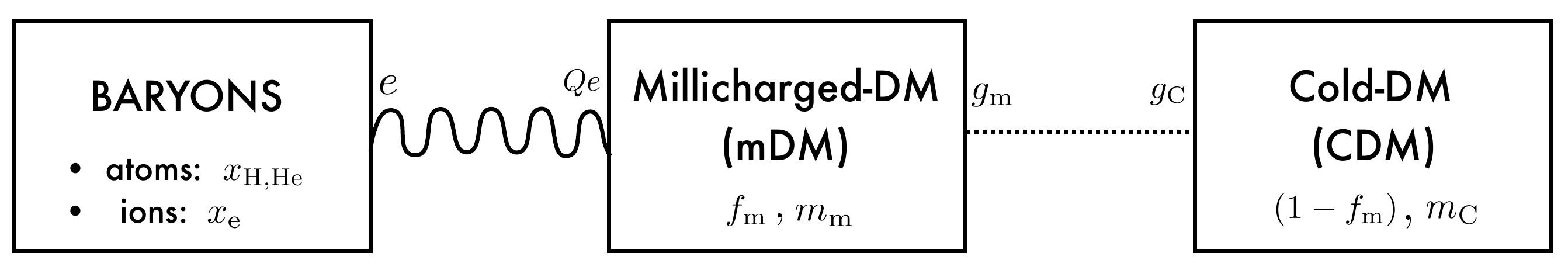}
\caption{ The structure of the dark sector studied in this paper. A fraction of millicharged DM (mDM), $f_\text{m}$, which carries electromagnetic charge $Q$, interacts with the baryonic bath which is composed of hydrogen (H), helium (He) with fraction $x_\text{H}$ and $x_{\text{He}}$ respectively and an ionized fraction $x_e$ composed of e$^-$, H$^+$. The latter goes from being order one at CMB to $\simeq 10^{-4}$ at $z\simeq 17$. The mDM also interacts with the rest of the cold DM (CDM) through a new dark long-range force, which couples with strength $g_\text{m}$ to mDM and $g_\text{C}$ to CDM. The same long-range force induces CDM self-interactions.
 }
\label{fig:cartoon}
\end{figure*}

In this paper, we revisit the possibility of baryon cooling by  a millicharged component of DM. We show that if mDM also has an additional long-range interaction with the rest of the dark matter (which forms a cold dark matter (CDM) bath), it may remain cold for longer, thereby greatly increasing the effective heat capacity of the hidden sector. This substantially improves the efficiency of baryon cooling, extending the region of parameter space in which we expect large effects on 21-cm cosmology. In particular, the setup, which we depict in Fig.~\ref{fig:cartoon}, can explain the recent  sky-averaged 21-cm spectrum measurement presented by the EDGES collaboration~\cite{Bowman2018}, while opening up the higher mass and smaller density regions of the mDM parameter space. This can be accomplished while ensuring that the momentum transfer between mDM and CDM remains small enough to circumvent the stringent CMB constraints.

The allowed parameter space in this scenario leads to precise predictions that will be tested in upcoming direct detection and collider experiments. The large heat capacity needed to cool the gas implies an upper bound of several GeVs on the CDM mass (see Fig.~\ref{fig:money_CDM_vector}), 
while BBN constraints require it to be larger than roughly $10\text{ MeV}$ \cite{Boehm:2013jpa}. Interactions between CDM and SM particles are radiatively generated  by loops of mDM particles, and the corresponding CDM-baryon cross section lies below the current direct detection limit from Xenon10 and SENSEI~\cite{Essig:2012yx,Essig:2017kqs,Abramoff:2019dfb}, but well within the reach of future runs of low threshold direct detection experiments like SENSEI and DAMIC~\cite{Battaglieri:2017aum}. Interesting constraints on the nature of CDM self-interactions can be inferred from colliding (sub-)clusters~\cite{Feng:2009mn,Kahlhoefer:2013dca} and from cluster ellipticity~\cite{Barrena:2002dp,MiraldaEscude:2000qt}, even though the reliability of the latter has been questioned~\cite{Peter:2012jh,Tulin:2017ara,Agrawal:2016quu}. The allowed parameter space for the mDM component in our new framework is greatly enlarged compared to the standard scenario, leading to a rich phenomenology that can motivate many upcoming beam dump and direct detection experiments. The mDM fraction, $f_\text{m}$, should be within $10^{-8}\lesssim f_\text{m}\lesssim 0.004$, where the upper bound comes from present CMB constraints~\cite{Dubovsky:2003yn,Dolgov:2013una,Kovetz:2018zan,dePutter:2018xte} and the lower bound is the minimal fraction required to explain EDGES while still being consistent with constraints from collider and beam dump experiments.
The mDM electric charge lies in the range $10^{-5}\lesssim Q\lesssim 1$,  such that the interactions between the mDM component 
and the baryons are strong enough to lead to sufficient cooling, while still being consistent with collider searches.
The mDM mass, $m_\text{m}$, is found to be in the range $10\text{ MeV}\lesssim m_\text{m}\lesssim 200\text{ GeV}$, 
where the lower bound again stems from BBN constraints\footnote{For the cross sections we consider here, both CDM and mDM are in thermal equilibrium with the SM in the early universe. In the simplest setup, the freeze out of CDM and mDM into light dark radiation can be in tension with $N_{\text{eff}}$ measurements at CMB and BBN posing a more stringent lower bound on both the mDM and CDM masses (see for example~\cite{Ackerman:mha,Vogel:2013raa}). This constraint can be circumvented easily for the mDM coupling to the SM by directly assigning a charge $Q$ under the SM electromagnetism. For the CDM-mDM mediator, the $N_{\text{eff}}$ bounds can be easily circumvented in non-minimal models and are ignored in the following.}, while the upper bound is derived by requiring that cooling is efficient enough to explain EDGES while not violating CMB constraints.

A large array of future experiments can potentially discover the mDM component in our scenario. The large-$Q$ region is constrained by present collider searches~\cite{Prinz:1998ua,Akers:1995az,Davidson:2000hf,Haas:2014dda} and will be further probed by a combination of future collider experiments such as milliQan~\cite{Haas:2014dda,Ball:2016zrp}, MicroBooNE~\cite{Magill:2018tbb}, LDMX~\cite{Berlin:2018bsc} and dedicated detector proposals in the NuMI or DUNE beamline~\cite{Kelly:2018brz,Harnik:2019zee}. Ground-based direct detection of mDM is challenging, due to mDM energy losses in the atmosphere, in the Earth's crust and in the shielding material of a given experiment~\cite{Starkman:1990nj,Mahdawi:2017utm}. For this reason, the best sensitivity to mDM in this coupling regime may be achieved by surface runs of low threshold detectors (e.g.\ Ref.~\cite{Crisler:2018gci}), which can detect the soft tail of the DM velocity distribution after atmospheric scatterings (see Ref.~\cite{Emken:2019tni} for an updated reach of these experiments for mDM). Pushing the reach to higher $Q$ can be achieved by either lowering the threshold of the detectors~\cite{Essig:2016crl,Schutz:2016tid,Knapen:2016cue,Budnik:2017sbu,Hochberg:2017wce,Bunting:2017net,Knapen:2017ekk,Rajendran:2017ynw,Griffin:2018bjn,Pospelov:2019vuf,Essig:2019kfe} or reducing the atmospheric shielding using balloon- and satellite-borne detectors~\cite{Emken:2019tni}.

The paper is organized as follows. In Sec.~\ref{eq:EDGESstd}, we summarize the relevant 21-cm cosmology formalism and  discuss the EDGES measurement. In Sec.~\ref{sec:mDM_std}, we explain why it is difficult for an mDM component with no interaction with the dominant CDM component to cool baryons efficiently.
Along the way, we develop an analytic understanding of the parametric behaviour of DM cooling after recombination, which qualitatively explains most of the features of the full numerical treatment, based on Ref.~\cite{Liu:2019bbm}. In Sec.~\ref{sec:bath}, we discuss quantitatively the advantages of the setup summarized in Fig.~\ref{fig:cartoon}. The allowed parameter space is derived by fitting the EDGES signal and accounting for other cosmological constraints. Sec.~\ref{sec:pheno} is dedicated to the phenomenological predictions of our setup and the future prospects for excluding or discovering both the mDM and the CDM components. Sec.~\ref{sec:conclusions} contains our conclusions. 

This paper is supplemented by four appendices in which we summarize many useful results that are at the core of both our analytic and numerical treatment.  App.~\ref{sec:21cm} summarizes a derivation of the thermodynamic equations for a system of two baths, first derived in Refs.~\cite{Tashiro:2014tsa,Munoz:2015bca}. The relevant rates are formulated in a model independent fashion in terms of the DM-baryon interaction form factors as usually done in the direct detection literature~\cite{Essig:2011nj}.  In App.~\ref{sec:study_case}, we compute the relevant form factors in the Born approximation. We discuss in particular the ones necessary to include for  DM scattering with hydrogen and helium atoms in 21-cm cosmology. We also assess the regions of parameter space in which the Born approximation breaks down, leaving a full quantum mechanical computation of the rates for future work. In App.~\ref{sec:analytical}, we discuss more quantitative details about the implementation of the framework described in  Fig.~\ref{fig:cartoon}. We discuss the  thermodynamic equations for a system of three baths and show how the standard mDM scenario previously discussed in Refs.~\cite{Munoz:2018pzp,Barkana:2018cct,Berlin:2018sjs} is parametrically extended as a function of the coupling between mDM and CDM. We then discuss in more detail the cosmological constraints from CMB on our scenario, relying on Refs.~\cite{Dubovsky:2003yn,Dolgov:2013una,Kovetz:2018zan,dePutter:2018xte,Boddy:2018wzy}. Lastly, App.~\ref{sec:bathmodels} discusses concrete models realizing the scenario of Fig.~\ref{fig:cartoon} which are used to derive the CDM component direct detection prospects.

\section{THE EDGES Observation
}\label{eq:EDGESstd}

The brightness temperature of the 21-cm hydrogen line, $T_{21}$, measures the contrast between redshifted 21-cm radiation and the CMB blackbody spectrum at the same frequency. As a function of redshift, this can be written as~\cite{Madau:1996cs}
\begin{equation}
\label{eq:Tbright}
T_{21} = \frac{1}{1+z} \left(T_s-T_\gamma \right)\left(1-e^{-\tau(T_s)}\right)\,,
\end{equation}
where $T_\gamma$ is the CMB temperature\footnote{Throughout this paper we assume that $T_\gamma$ is given by the CMB temperature. See Refs.~\cite{Feng:2018rje,Ewall-Wice:2018bzf,Pospelov:2018kdh} for examples in which  this assumption is relaxed, giving rise to important alterations to the 21-cm global signal.},  $T_s$ is the spin temperature of the 21-cm line and $\tau(T_s)$ its optical depth. The latter two quantities are defined as  
\begin{alignat}{1}
&\frac{n_1}{n_0} \equiv \frac{g_1}{g_0} e^{-E_{21}/T_s}\simeq 3\left(1-\frac{E_{21}}{T_s}\right)\ ,\\
&\tau(T_s) \simeq\frac{3 \lambda_{21}^2A_{10}n_{\rm H}}{16T_s H}\, ,\label{eq:opticaldepth}
\end{alignat}    
where $H$ is the Hubble parameter at a given redshift, $\lambda_{21}=1/E_{21}\simeq 21\text{ cm}$ and $A_{10}\simeq 2.9\times10^{-15}{\rm\ s^{-1}}$ is the Einstein spontaneous emission coefficient for the 21-cm transition. The spin temperature $T_s$ characterizes the ratio between the number densities of the triplet and singlet hyperfine sub-levels of the hydrogen 1s state. When $T_s$ is larger (smaller) than $T_\gamma$, we expect to see an emission (absorption) feature in $T_{21}$. 

The EDGES experiment claimed the first measurement of the brightness temperature at the cosmic dawn ($z\simeq 17$)~\cite{Bowman2018}. During this epoch, the first stars begin to emit Ly-$\alpha$ radiation, affecting the spin temperature as Ly-$\alpha$ photons pass through hydrogen clouds by driving spin-flip transitions between the hyperfine sub-levels via the Wouthuysen-Field effect \cite{1952AJ.....57R..31W,1958PIRE...46..240F}. The rate of these transitions quickly become much faster than downward transitions through stimulated emission by CMB photons, driving the temperature of the baryons $T_\text{b}$, the spin temperature $T_s$ and the temperature of the Ly-$\alpha$ spectrum to a common value.  

One can thus quite generally expect $T_\text{b} \lesssim T_s \lesssim T_\gamma$, with the ratio $T_s/T_\text{b}$ depending on the strength of the Wouthuysen-Field effect\footnote{Our computation of $T_s$ (and the subsequent determination of $T_{21}$ in Eq.~\eqref{eq:t21}) neglects astrophysical effects that could lead to extra heating of the gas, diminishing the absorption feature at cosmic dawn. Even under the assumption that X-ray heating is turned off at $z\simeq 17$ \cite{Bowman2018} (see \cite{Madau:1996cs,Chen:2003gc,Cohen:2016jbh} for a quantitative discussion of X-ray heating), a full self consistent computation should include the hydrogen heating from its scattering with the Ly-$\alpha$ photons \cite{Chen:2003gc} and the CMB heating first computed in Ref.~\cite{Venumadhav:2018uwn}. The latter has been shown to be the dominant effect leading to up to $50\%$ deviations in the 21-cm brightness temperature. Since the magnitude of these effects is not enough to invalidate our picture, we defer a careful study of these, as well as of the shape of the absorption peak, for future studies.}. The {strongest} absorption peak in the brightness temperature defined in Eq.~\eqref{eq:Tbright} is  obtained by saturating the inequality on the left ($T_s=T_{\text{b}}$), giving
\begin{equation}
T_{21}\vert_{z=17}=  0.35\text{ K} \left(1-\left.\frac{T_\gamma}{T_{\text{b}}}\right|_{z=17}\right)\label{eq:t21}\,.
\end{equation}

The $\Lambda\text{CDM}$ prediction for $T_\text{b}$ and $T_\gamma$ at $z=17$ gives $T_{21}^{\Lambda\text{CDM}}\vert_{z=17}= -0.22 \text{ K}$. This lower bound is in $3.8\sigma$ tension with the central value of the EDGES measurement~\cite{Bowman2018}, whose central value and 99\% confidence intervals are
\begin{equation}
T_{21}^{\text{EDGES}}\vert_{z=17}=-0.5^{+0.2}_{-0.5}\text{ K}\ .\label{eq:edges}
\end{equation}

This observed tension, if it persists, would point to a deviation from standard cosmology, and has drawn significant  attention. Previous work in the literature has focused on two possible solutions: (i) either 21-cm photons at $z=17$ were hotter than the CMB due to astrophysical sources \cite{Feng:2018rje,Ewall-Wice:2018bzf,Fialkov:2019vnb} or new physics~\cite{Pospelov:2018kdh}, or (ii) the gas temperature at $z\simeq17$ was colder than the one predicted by $\Lambda\text{CDM}$ due to cooling induced by the scattering of baryons with some fraction of the cold DM bath~\cite{Tashiro:2014tsa,Munoz:2015bca,Munoz:2018pzp,Barkana2018,Barkana:2018cct,Berlin:2018sjs,DAmico:2018sxd,Liu:2018uzy}. In this paper we continue to investigate the latter option.

Using Eqs.~\eqref{eq:t21} and~\eqref{eq:edges} and assuming that $T_\gamma$ is simply the CMB temperature, the gas temperature inferred from the EDGES measurement is 
\begin{equation}
T_{\text{b}}^{\text{EDGES}}\vert_{z=17}=3.2_{-1.55}^{+1.9}\text{ K} \,, 
\end{equation}
with the deviation from the $\Lambda\text{CDM}$ prediction being given by
\begin{equation}
\Delta T_{\text{b}}^{\text{EDGES}}\equiv (T_{\text{b}}^{\text{EDGES}}-T_{\text{b}}^{\Lambda\text{CDM}})\vert_{z=17}=-3.6^{+1.9}_{-1.55}\text{ K}\, .\label{eq:temperatureEDGES}
\end{equation}
The DM bath must therefore cool the baryons by \emph{at least} \SI{1.7}{K} to a temperature of \SI{5.1}{K} in order to accommodate the EDGES result within the 99\% CL. This is the amount of cooling that will be conservatively required throughout this paper. 

Before leaving this section it is worth pointing out that the 21-cm global signal measurement are experimentally intricate and face significant background subtraction challenges. For this reason, it has been suggested that the significance of the EDGES result could be appreciably reduced depending on the background subtraction procedure~\cite{Hills:2018vyr,Tauscher:2018uxi,Bradley:2018eev,Spinelli:2019oqm}. Future and ongoing experiments are likely to provide more insights and possibly independent tests on the EDGES result \cite{Voytek:2013nua,Singh:2017cnp,2018MNRAS.478.4193P,2019JAI.....850004P,Burns:2017ndd,Nhan:2018cvm}. Nevertheless, we believe it is an important theoretical question to classify the classes of dark sectors that could give signals in 21-cm cosmology without being already excluded by other complementary probes. In the remainder of the paper, we will explore this question, focusing on DM cooling the baryons through elastic collisions.

\section{Standard Millicharged Dark Matter} 
\label{sec:mDM_std}

Having presented  the EDGES result, let us now discuss the presence  of an mDM component that interacts with the baryonic gas. We shall argue that such an addition to $\Lambda\text{CDM}$ is insufficient to explain the EDGES result consistently with the existing cosmological bounds.

We begin by estimating the mDM-baryon interaction rate required for the mDM to account for the necessary gas cooling presented above in Eq.~\eqref{eq:temperatureEDGES}. To do so we study the evolution of the gas temperature $T_\text{b}$, given by\footnote{Eq.~\eqref{eq:Tbaryon} neglects for simplicity heating from both the DM-baryon bulk relative motion and the Compton scattering with CMB photons.  Both of these effects are included in our numerical treatment, which is based on Ref.~\cite{Liu:2019bbm}. More details about this are given in Appendix \ref{sec:analytical}.}  (see App.~\ref{sec:21cm} for a detailed derivation and the full expression)
\begin{equation}
\dot{T}_{\text{b}}\simeq-2HT_{\text{b}}- \frac{2 x_\text{b} \mu_{\text{bm}}}{m_\text{b}+m_{\text{m}}}\frac{\langle \Gamma_{\rm b} v_{\text{rel}}^2\rangle}{\langle v_{\text{rel}}^2\rangle}(T_\text{b}-T_{\text{m}})\, . \label{eq:Tbaryon}
\end{equation}
The first term on the right hand side accounts for adiabatic cooling due to the expanding universe while the second term describes the cooling of the gas due to the mDM-baryon interactions. $T_\text{m}$ is the mDM temperature, $x_\text{b}$ is the fraction of  baryons interacting with mDM, $\mu_{\text{bm}}$ is the reduced mass of the interacting baryon and the mDM, and 
\begin{equation}
\label{eq:Gammab}
\Gamma_{\text{b}}=f_\text{m} \frac{\rho_{\rm DM}}{m_{\rm m}}\sigma_T^{\text{bm}} v_{\text{rel}}
\end{equation}
is the mDM-baryon interaction rate, with $\sigma_T^{\rm bm}$ the transfer cross section (see Eqs.~\eqref{eq:mDM} and~\eqref{eq:deftransf}) and $v_{\rm rel}$ the relative velocity between the colliding mDM and baryon particles.  We define  $f_{\text{m}}\equiv \rho_{\rm mDM}/\rho_{\rm DM}$ to be the energy density fraction of the mDM. The symbol $\langle\dots\rangle$ indicates thermal averaging over the velocity distributions of both fluids.

We calculate the mDM-baryon transfer cross section assuming that the Coulomb potential is regulated by the Debye thermal mass of the photon, $m_\gamma^2\simeq \alpha_{\rm EM}n_e/T_{\rm b}$. Since (i) $m_\gamma$ is much smaller than the typical momentum transfer $q\simeq \mu_{\rm bm} v_{\rm rel}$, and (ii) for the relevant charges $Q$ considered here, $m_\gamma\ll \mu_{\rm bm}v_{\rm rel}^2/Q \alpha_{\rm EM}$, $\sigma_T^\text{bm}$ should be evaluated in the classical regime described in~\cite{Khrapak:2003kjw,1308514,Feng:2009hw,Tulin:2013teo}, that is,\
\begin{equation}
\sigma_T^{\text{bm}}\simeq\frac{2\pi Q^2 \alpha_{\text{EM}}^2}{\mu^2_{\text{m}} v_{\text{rel}}^4}\log\left(\frac{T_{\rm b}m_p\mu^2 v_{\text{rel}}^4}{Q^2\alpha_{\text{EM}}^3 \rho_{\rm b}}\right)\,.\label{eq:mDM}
\end{equation}
We stress that this expression has a different log factor compared to the one commonly used in the literature~\cite{McDermott:2010pa,Dvorkin:2013cea,Munoz:2018pzp}. The cross section in Eq.~(\ref{eq:mDM}) is evidently  enhanced as $v_{\text{rel}}^{-4}$ at low relative velocities.  Substituting this expression into Eq.~\eqref{eq:Tbaryon}, we see that the  effect of mDM-baryon scattering on the temperature evolution grows rapidly as redshift decreases.

In the absence of the second term on the RHS of Eq.~\eqref{eq:Tbaryon}, baryons cool only adiabatically.  This is insufficient (by an order one amount) for the temperature evolution to be consistent with the EDGES measurements.  Consequently, in order for significant cooling to occur, the rate of energy transfer must be at least as large as the energy loss from adiabatic expansion, but not much larger. We therefore require the cooling rate to be comparable to Hubble at around $z=17$ in order to explain the EDGES discrepancy in Eq.~\eqref{eq:temperatureEDGES}. This gives 
\begin{equation}
\left.\langle v_{\rm rel}^2\rangle^{1/2}\sigma_T^{\rm bm}\right\vert_{z= 17}\simeq \frac{m_{\rm m}^2m_p}{x_{\rm b}f_{\rm m}\mu_{\rm m}^2 \rho_{\rm DM}}\left.\frac{H}{1-T_{\text{m}}/T_{\text{b}}}\right\vert_{z=17}\ ,\label{eq:rate}
\end{equation}
where a much smaller rate will not be able to cool the baryons, and a much larger one would cause an exponential decrease in the baryon temperature. 

The mDM temperature at $z=17$ must lie between 
\begin{equation}
\frac{m_{\text{m}}}{m_p}\frac{\Omega_{\text{b}}}{f_{\text{m}}\Omega_{\text{DM}}}\vert\Delta T^{\rm EDGES}_{\text{b}}\vert\lesssim T_{\text{m}}< T_{\text{b}}\ ,\label{eq:temp}
\end{equation}
where $\Omega_\text{b}/\Omega_{\text{DM}}\simeq0.18 $. 
The upper bound on $T_{\text{m}}$ ensures that the mDM bath can cool the gas, while the approximate lower bound can be derived by assuming that the additional heat per unit volume removed from the baryons is transferred to a zero temperature mDM fluid, so that $n_\text{b} |\Delta T_\text{b}^\text{EDGES}| \sim n_\text{m} T_\text{m}$.

\begin{figure*}[t]
\centering
\includegraphics[width=0.9\textwidth]{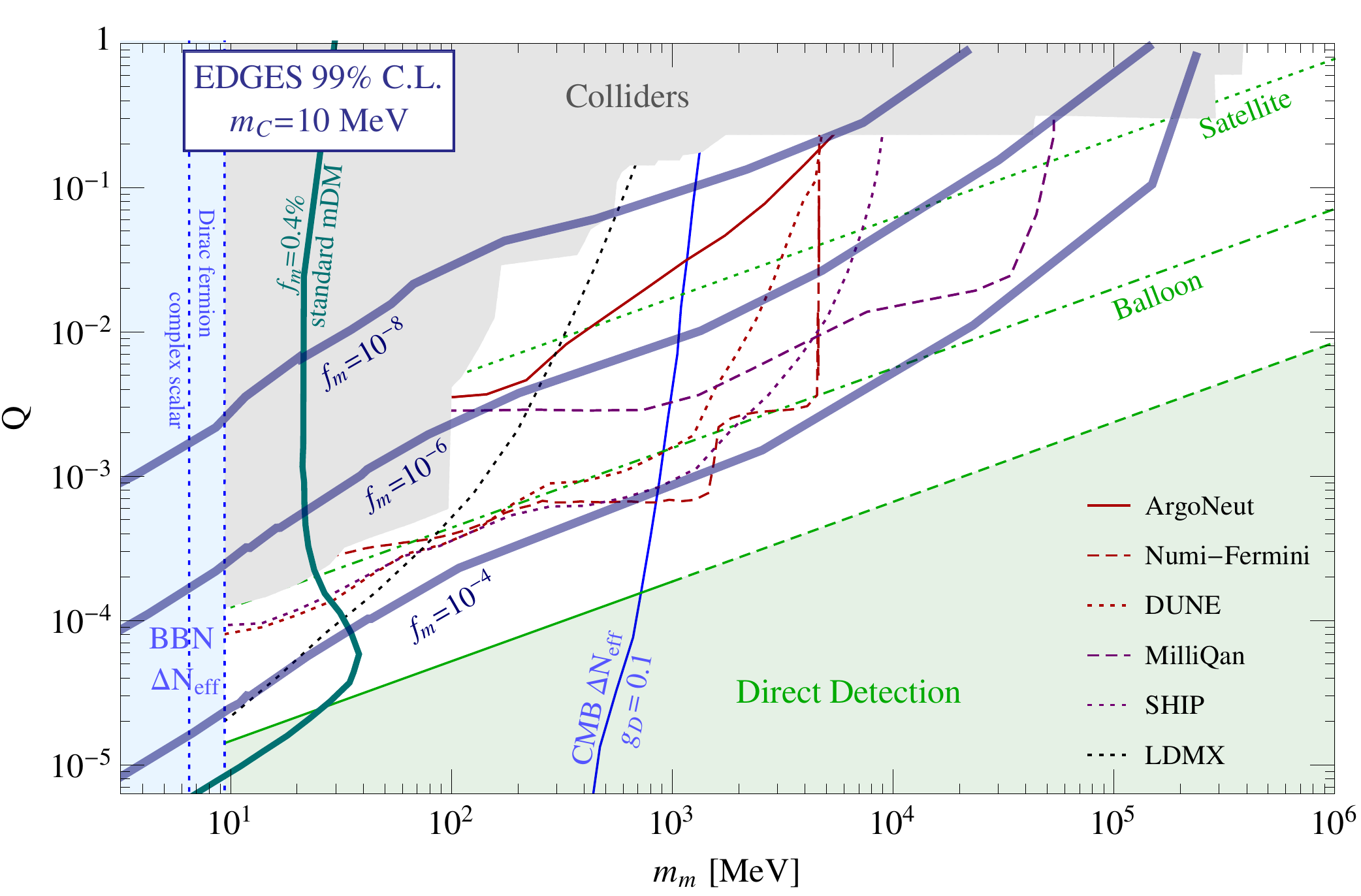}
\caption{Parameter space of the scenario described in Fig.~\ref{fig:cartoon} in the plane ($m_{\text{m}}$, $Q$) where we fix $m_{\text{C}}=10\text{ MeV}$ to maximize the heat capacity of the CDM bath and the maximal $\alpha_\text{m} \alpha_\text{C} $ allowed by CMB bounds \cite{Dubovsky:2003yn,Dolgov:2013una,Kovetz:2018zan,dePutter:2018xte,Boddy:2018wzy}. The \textbf{dark blue} contours give the mDM fraction $f_{\text{m}}$ required for a given $(m_{m}, Q )$ point to fit the upper value of the 99\% CL interval of the EDGES measurement in the setup of Fig.~\ref{fig:cartoon}. For a fixed $f_{\text{m}}$ the entire region above the dark blue contour can be probed by reducing $\alpha_{\text{m}}\alpha_{\text{C}}$ (see text for details). For comparison, the \textbf{dark green} contour shows the standard mDM case where $0.4\%$ of mDM alone provides the baryonic cooling. The \textbf{light blue} region for $m_\text{m}<10\text{ MeV}$ is robustly excluded by BBN contraints on $N_{\text{eff}}$~\cite{Munoz:2018pzp,Barkana:2018cct,Berlin:2018sjs,Creque-Sarbinowski:2019mcm}, 
the two dotted lines distinguish between the case in which mDM is a scalar or a Dirac fermion. The \textbf{gray shaded} area is a collection of different constraints taken from Refs.~\cite{Davidson:2000hf,Badertscher:2006fm}, plus limits on millicharge particles from milliQ at SLAC~\cite{Prinz:1998ua}, searches for  low ionizing particles in CMS at the LHC~\cite{CMS:2012xi} and the new constraints from LSND and MiniBooNE derived in Ref.~\cite{Magill:2018tbb}. The region on the left of the \textbf{blue line} is excluded by CMB constraints on $N_{\text{eff}}$ only when mDM couples to a dark photon with coupling $g_D=0.1$. The \textbf{green region} is excluded by present direct detection experiments as shown in Ref.~\cite{Emken:2019tni}. The \textbf{green dashed line} indicates our extrapolation of the results in Ref.~\cite{Emken:2019tni} to higher masses (see discussion in the text). The \textbf{red/black/magenta lines} indicate the Fermilab/SLAC/CERN effort to probe mDM. \textbf{Solid/dashed/dotted lines} give a rough sense of the short/medium/long time scale of the proposal. \textbf{Solid red} is the ArgoNeut sensitivity derived in Ref.~\cite{Harnik:2019zee}, \textbf{dashed red} is the sensitivity of the Fermini proposal at NuMI~\cite{Kelly:2018brz} (see Ref.~\cite{Harnik:2019zee} for a more conservative reach based on ArgoNeut at NuMI), \textbf{dotted red} is the DUNE reach~\cite{Magill:2018tbb} while \textbf{dotted black} is the LDMX reach~\cite{Berlin:2018bsc}. \textbf{Dashed magenta} is the milliQan reach as \cite{Haas:2014dda} while \textbf{dotted magenta} is the SHiP sensitivity~\cite{Magill:2018tbb}. The \textbf{dash-dotted/dotted green lines} indicate the reach of a SENSEI-like dark matter detector on a balloon/satellite with 0.1 gram-month exposure~\cite{Emken:2019tni}.} \label{fig:money_milli}
\end{figure*}

Using the lower bound for $T_{\text{m}}$ in Eq.~\eqref{eq:rate}, we get an estimate of the smallest rate required to explain the EDGES measurement. In the limiting case where $T_{\text{m}}$ is exactly equal to $T_{\text{b}}$ in Eq.~\eqref{eq:temp}, we find
\begin{equation}
m_{\text{m}}\lesssim \SI{67}{\mega\eV} \left( \frac{f_{\text{m}}}{0.4\%} \right) \,. \label{eq:upperboundmDM}
\end{equation}
The upper bound on $m_\text{m}$ occurs at the point where the number density of mDM is too small to absorb the required amount of heat from the baryons. CMB constraints for a typical mDM charge required for sufficient cooling exclude $f_{\text{m}}\gtrsim 0.4\%$~\cite{Dvorkin:2013cea,Kovetz:2018zan,Boddy:2018wzy}. Fig.~\ref{fig:money_milli} shows in dark green the value of $Q$ as a function of $m_\text{m}$ required to fit the EDGES signal with $f_{\text{m}}=0.4\%$, computed using a full numerical treatment. The turnaround is qualitatively explained by Eq.~\eqref{eq:upperboundmDM}: at higher mDM masses the heat capacity of the mDM bath becomes insufficient to cool the baryonic gas. The full numerical treatment includes the heating of the baryons from Compton scattering with CMB photons, explaining why the turnover in Fig.~\ref{fig:money_milli} happens at a slightly smaller mDM mass than predicted by Eq.~\eqref{eq:upperboundmDM}.

Given the upper bound in Eq.~\eqref{eq:upperboundmDM} and the required couplings to baryons in Eq.~\eqref{eq:rate}, the range of possible $m_\text{m}$ and $f_\text{m}$ that can explain the EDGES result is severely restricted by constraints on the effective number of relativistic degrees of freedom from BBN and CMB. 
These are notoriously  difficult to accommodate given a particle freezing out at temperatures below the QCD phase transition. Even though some caveats to this argument can be constructed (see Ref.~\cite{Berlin:2018sjs} for a discussion), these difficulties substantially reduce the appeal of this minimal mDM scenario as an explanation of the  EDGES result.

\section{Millicharged Dark Matter Interacting With a Cold Dark Matter Bath}
\label{sec:bath}

In the remainder of this paper, we demonstrate how a new long-range interaction between a small fraction of mDM and the rest of the DM bath significantly improves upon the standard mDM scenario in explaining the EDGES observation. We first begin with an overview of our framework in Sec.~\ref{sec:framework}. We then explain how the mDM parameter space is extended in Sec.~\ref{sec:scatteringparam}, discuss the cosmological constraints on the dark sector in Sec.~\ref{sec:CMB}, and finally describe the importance of mDM-neutral atom scattering in our setup in Sec.~\ref{sec:atoms}. A more detailed discussion of each of these aspects of our model is included in the appendices.

\subsection{The Framework}
\label{sec:framework}
Before diving into a detailed analysis, we wish to introduce the main features of our new framework, illustrated in Fig.~\ref{fig:cartoon}, and summarize our main results.  The dark sector is composed of a CDM and an mDM component. The mDM, of mass $m_{\rm m}$, constitutes a fraction $f_\text{m}$ of the total DM energy density. The mDM-baryon long-range interaction is controlled by the mDM charge $Q$, which may or may not stem from the presence of a new light mediator. 
The novelty in our setup is that the mDM fraction also interacts with the remaining CDM component, of mass $m_\text{C}$, through a distinct  long-range hidden interaction controlled by the coupling $g_\text{m} g_\text{C}$. The same interaction also induces  a CDM self-interaction proportional to $g_\text{C}^2$. The two long-range interactions of our setup imply the existence of one or two new light mediators with masses below a keV, which is the typical size of the exchange momentum in scattering collisions during the cosmic dawn. 
  
The long-range force between mDM and CDM opens up the mDM parameter space at higher masses (up to $m_\text{m}\lesssim 200\text{ GeV}$)  and smaller dark matter fraction (down to $f_\text{m}\gtrsim 10^{-8}$).  This is because the cooling is now driven by the CDM bath, with the mDM acting as a mediator between CDM and the baryons. As we show in Sec.~\ref{sec:scatteringparam}, the CDM mass $m_\text{C}$ must lie below a few GeV in order to have a large enough heat capacity to cool the gas sufficiently.

The allowed parameter space of our framework is mostly determined by ensuring that the mDM-baryon and mDM-CDM couplings are consistent with CMB constraints, as discussed in Sec.~\ref{sec:CMB}. In Fig.~\ref{fig:money_milli}, we show three contours on the $m_\text{m}$ -- $Q$ plane where sufficient cooling of the baryonic bath is achieved in our framework in order to explain the EDGES result for $f_\text{m} = 10^{-4}$, $10^{-6}$ and $10^{-8}$. We have fixed $m_\text{C} = \SI{10}{\mega\eV}$ and $g_\text{m}g_\text{C}$; these two parameters can vary over a broad range of values without affecting our qualitative results, as we will explain in greater detail in Sec.~\ref{sec:CMB} below.  
The white region of Fig.~\ref{fig:money_milli} represents the available mDM parameter region, bounded between present collider bounds~\cite{Davidson:2000hf,Badertscher:2006fm,Prinz:1998ua,CMS:2012xi,Magill:2018tbb} and direct detection bounds~\cite{Emken:2019tni}. We also show the prospects of exploring this region of parameter space with upcoming experiments. This, together with the direct detection prospects for the CDM component, which has a loop-induced interaction with the SM, will be discussed in Sec.~\ref{sec:pheno}.

\subsection{Cooling Parametrics}
\label{sec:scatteringparam}
The long-range interaction between  CDM and mDM induces a Rutherford-like transfer cross section between the two baths given by
\begin{equation}
\sigma_T^{\text{mC}}=\frac{2\pi  \alpha_{\text{C}}\alpha_{\text{m}}}{\mu^2_{\text{mC}} v_{\text{rel}}^4}\log\left(\frac{\mu_\text{mC}^2 v_{\text{rel}}^4}{{\alpha_\text{m}} \alpha_\text{C} m_\phi^2}\right)\label{eq:CDM}\ ,
\end{equation}
where we defined $\alpha_\text{m}=g_\text{m}^2/4\pi$, $\alpha_\text{C}=g_\text{C}^2/4\pi$ and again we assume the mediator mass $m_\phi$ satisfies $m_\phi\ll \mu_{\rm mC}v_{\rm rel}^2/\sqrt{\alpha_{\rm m} \alpha_{\rm C}}$ as in Eq.~\eqref{eq:mDM}. The exact nature of the mediator (scalar or vector) will not modify the discussion here. The explicit example of a vector mediator will be discussed in both Sec.~\ref{sec:pheno} and Appendix~\ref{sec:bathmodels}. 

As in the standard mDM case, due to the strong velocity dependence, the cooling of baryons is largest at low redshifts, where the velocity of the baryonic bath is the smallest. The crucial difference is that in our framework, the cooling is achieved dominantly by an mDM fluid that is coupled to the CDM bath. In a manner similar to  Eq.~\eqref{eq:rate}, sufficient cooling of baryons to explain the EDGES measurement is approximately achieved when
\begin{equation}
	\left.\langle v_{\rm rel}^2\rangle^{1/2}\sigma_T^{\rm bm}\right\vert_{z= 17}\simeq \frac{m_{\rm m}^2m_p}{x_{\rm b}f_{\rm m}\mu_{\rm m}^2 \rho_{\rm DM}}\left.\frac{H}{1-T_{\text{C}}/T_{\text{b}}}\right\vert_{z=17}\,, 	
\end{equation} 
with the important difference from Eq.~\eqref{eq:rate} being that the rate now depends on $T_{\rm C}/T_{\rm b}$, where $T_\text{C}$ is the CDM temperature, instead of $T_{\rm m}/T_{\rm b}$.

Following the derivation of Eq.~\eqref{eq:temp} closely, we  find that the CDM temperature at $z=17$ is now bounded by 
\begin{equation}
\frac{m_{\text{C}}}{m_p}\frac{\Omega_{\text{b}}}{\Omega_{\text{C}}}\vert\Delta T^{\rm EDGES}_{\text{b}}\vert\lesssim T_{\text{C}}< T_{\text{b}}\ ,\label{eq:temp2}
\end{equation}
where the lower bound is set by the minimal amount of heat that CDM absorbs from cooling the baryons through the mDM fraction. Saturating the inequalities  above results in an upper bound on $m_{\rm C}$,
\begin{equation}
m_{\text{C}}\lesssim 18\text{ GeV}\ ,\label{eq:upperbound}
\end{equation}
where the gain compared to the standard mDM bound in Eq.~\eqref{eq:upperboundmDM} is given by the fact that the heat capacity of the dark sector is now set by the number density of CDM, which is much larger than that of the mDM component. In deriving the approximate upper bound in Eq.~(\ref{eq:upperbound}), we have neglected Compton heating from the CMB photons. For $m_\text{C}$ values well below the bound in Eq.~(\ref{eq:upperbound}), the heat capacity of the CDM is always large enough to absorb heat from the mDM fluid efficiently, and as a result the contours shown in Fig.~\ref{fig:money_milli} are insensitive to the exact value of $m_\text{C}$.

\begin{figure*}[t]
\centering
\includegraphics[width=0.8\textwidth]{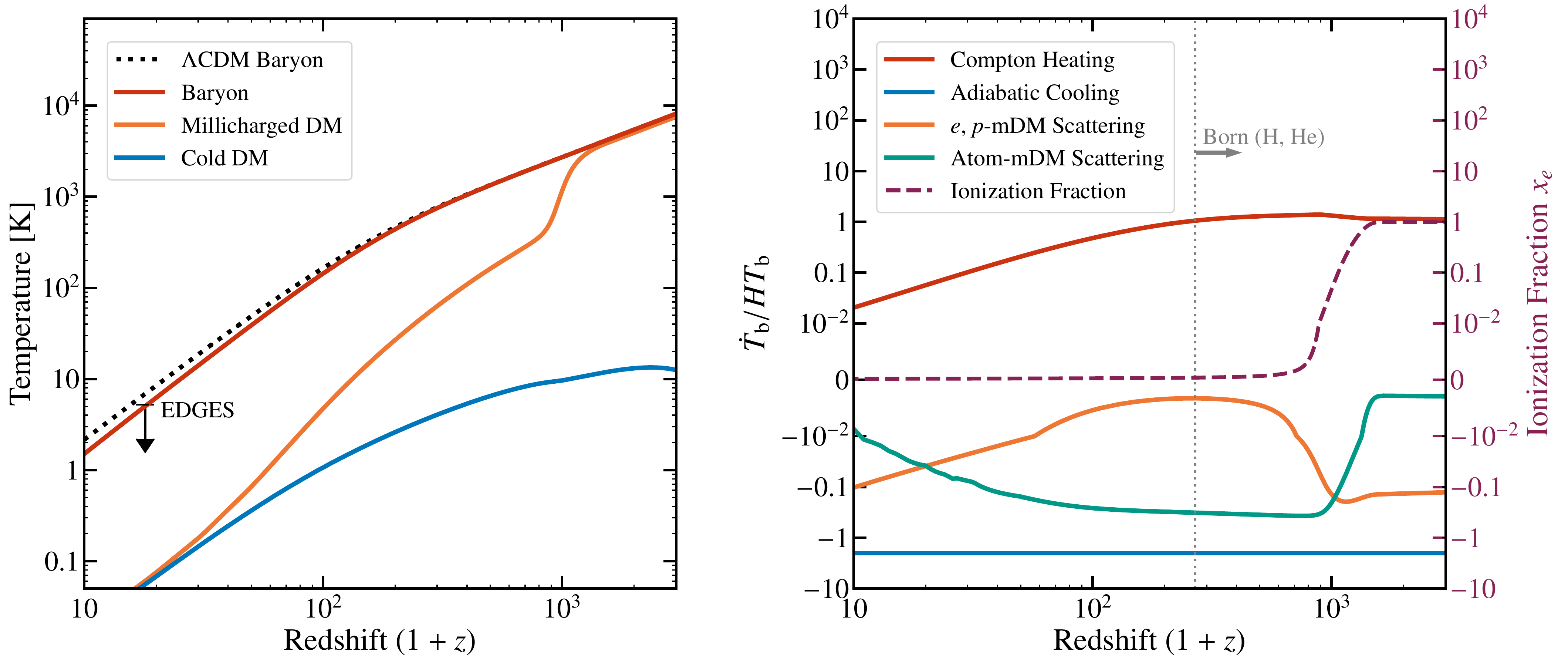}
\caption{Temperature evolution ({\bf left}) together with ionization evolution and heating/cooling rates $\dot{T}_\text{b}/HT_\text{b}$ of relevant processes ({\bf right}) for benchmark parameters given by $Q=6\times 10^{-4}$, $m_\text{m} = 2\text{ GeV}$ and $f_\text{m} = 10^{-4}$, with $m_\text{C}=100\text{ MeV}$ and $\alpha_\text{m} \alpha_\text{C}=4 \times 10^{-16}$ setting the interactions between mDM and CDM. At $z\simeq 3000$, the mDM temperature is close to the baryon temperature while the CDM bath is significantly colder. As the free electron fraction $x_e$ drops, the mDM decouples from the baryons, thermalizing with the CDM while cooling the baryonic gas. These parameters are consistent with known experimental constraints on $Q$, as well as CMB constraints on $g_\text{m} g_\text{C}$. They also produce sufficient cooling to be consistent with the upper limit of the EDGES 99\% CL ({\bf arrow, black}). For the range of mDM masses considered here, the cooling of the baryons from both scattering with the ionized fraction ({\bf solid, orange}) and with hydrogen and helium atoms ({\bf solid, green}) are important.  At low redshifts (i.e.\ low relative velocities), the Born approximation for scattering between mDM and atoms breaks down ({\bf gray, dotted}). In particular, the formation of mDM-baryon bound states could increase the cooling rate from mDM scattering with hydrogen and helium at lower redshift. Since this effect would only enlarge the parameter space presented here, we leave an improved calculation of this effect for future work. We also neglect the effect of the drag forces between the baths; this will modify the temperature evolution, introducing an $\mathcal{O}(1)$ change in the required mDM charge for sufficient cooling. This effect does not modify the qualitative features of our solution, but it is accounted for in Fig.~\ref{fig:money_milli}. A quantitative assessment of this effect can be found in Appendix~\ref{sub:three_fluids_system}.} 
\label{fig:temp_ev}
\end{figure*}

In Fig.~\ref{fig:temp_ev}, we show a typical temperature history in our framework where baryons are cooled sufficiently to give the EDGES result. The dark sector coupling $g_\text{m} g_\text{C}$ has been chosen so that the mDM bath is coupled to baryons at recombination; this is important for CMB constraints on our setup, which we will discuss below. For simplicity, in Fig.~\ref{fig:temp_ev} we also neglect the drag force between the mDM and CDM fluids, which  induces an extra heating source for the baryons at high redshift. These effects are included in our full numerical solution, and have a qualitative impact on some of our parametric estimates. A complete discussion of our full numerical treatment can be found in Appendix~\ref{sub:three_fluids_system}.

\subsection{Cosmological Bounds}
\label{sec:CMB}

Like the pure mDM model discussed in Sec.~\ref{sec:mDM_std}, this setup faces important cosmological constraints from the CMB power spectrum measurement (which constrains both the momentum transfer between baryons and CDM as well as $N_\text{eff}$), and from the primordial abundance of light nuclei produced during Big Bang nucleosynthesis (BBN).

As we have mentioned above, for $f_{\rm m}\lesssim 0.4\%$ the CMB power spectrum bounds on mDM vanishes, since it can be considered to be a component of the baryonic fluid whose contribution to the total baryon energy density $\Omega_{\rm b}$ lies within the uncertainty of its observed value. The addition of a new CDM-mDM interaction leads to a momentum drag between the baryonic and the CDM fluid, which is constrained by measurements of the CMB anisotropy power spectrum~\cite{Dubovsky:2003yn,Dolgov:2013una,Kovetz:2018zan,dePutter:2018xte}. While a dedicated analysis of our framework is beyond the scope of this paper, a simple recasting of existing constraints from Refs.~\cite{Dubovsky:2003yn,Dolgov:2013una,Kovetz:2018zan,dePutter:2018xte} can be performed as long as the mDM is tightly coupled to the baryons before recombination; this can only be achieved if the mDM coupling to the CDM bath is sufficiently weak (see App.~\ref{sub:constraints} for derivation):
\begin{equation}
	\alpha_\text{C} \alpha_\text{m} < 6.3 \times 10^{-7} Q^2 \left(\frac{m_\text{C}}{\SI{100}{\mega \eV}}\right) \,.
\end{equation}
Under the above assumption, the standard constraint for mDM interacting with the baryons can be rescaled to obtain an upper limit on the CDM-mDM interaction.  One finds (see App.~\ref{sub:constraints} for details) 
\begin{equation}
    \sigma_{T}^{\mathrm{mC}}V_\text{mC}^4 \lesssim \SI{7.8e-40}{\centi\meter\squared} \left( \frac{m_{\mathrm{C}}+m_{\mathrm{m}}}{m_{p}} \right) \left(\frac{0.4\%}{f_{\rm m}} \right)\, ,\label{eq:cmbconstrain}
\end{equation}
where $V_\text{mC}$ is the bulk relative velocity between the CDM and both the mDM bath and the baryons, which we take to be \SI{29}{\kilo\meter\per\second} before kinetic decoupling at $z = 1010$; this is roughly the rms value expected in $\Lambda$CDM~\cite{Ali-Haimoud:2013hpa,Tseliakhovich:2010bj}. Note that by virtue of Eq.~\eqref{eq:CDM}, the LHS depends only logarithmically on $V_\text{mC}$.  A more careful analysis of the CMB constraints on our scenario is expected to lead to weaker bounds on the mDM-CDM interaction strength.

Unlike the standard mDM scenario presented in Sec.~\ref{sec:mDM_std}, the CMB constraints can be relaxed by reducing $f_\text{m}$ and/or choosing $m_\text{m}+ m_{\text{C}}\gg m_p$: this is one of the key reasons why the range of viable mDM masses can be so large. In our numerical scans over $m_{\rm m},\,m_{\rm C}$ and $f_{\rm m}$  we choose the largest 
 possible coupling which controls the CDM-mDM cross section allowed by the CMB bound in Eq.~\eqref{eq:cmbconstrain}.  This choice of coupling also allows the mDM component to thermally couple to the CDM bath before $z\simeq 17$, which is important for efficient cooling of the baryons in most of the parameter space. Given the upper bound on $\alpha_{\text{m}}\alpha_{\text{C}}$ from the CMB constraint and the minimal required value to recouple the mDM bath back to the CDM bath at low redshift, we can explore the allowed range of $\alpha_{\text{m}}\alpha_{\text{C}}$ in the $m_\text{m}$ -- $Q$ plane, for a given CDM mass and mDM fraction. This is shown in  Fig.~\ref{fig:qeff} for $f_\text{m}=10^{-4}$ and $m_\text{C}=100\text{ MeV}$. The wide range of couplings over which we can explain the EDGES measurement  shows that our scenario does not rely on any special fine-tuning between $Q$ and $\alpha_{\text{m}}\alpha_{\text{C}}$ in most of the parameter space. 

BBN and CMB constraints on $N_\text{eff}$ set a lower bound on the masses of both mDM and CDM. For the couplings under consideration, both mDM and CDM are in thermal equilibrium with the SM at high redshift. Moreover, mDM and CDM annihilations into the light states mediating the long-range interactions between mDM and the baryons in Eq.~\eqref{eq:mDM} and between CDM and mDM in Eq.~\eqref{eq:CDM} would inject new relativistic degrees of freedom, which are highly constrained by $N_{\text{eff}}$ limits set by the CMB and the primordial abundance of nuclei.

To avoid any modifications to BBN, we take both masses to be above 10 MeV. For mDM, the value of $N_\text{eff}$ resulting from annihilation into dark photons has been studied in detail in Ref.~\cite{Vogel:2013raa}, leading to a stronger lower bound on the mDM mass ($m_\text{m}\gtrsim 400\text{ MeV}$) shown in Fig.~\ref{fig:money_milli} as a light blue shaded region. This bound assumes $g_D=0.1$ where the mDM charge is given by $Q=g_D \epsilon/g_\text{EM}$ and $\epsilon$ is the mixing between the dark photon and the SM photon (see also footnote \ref{eq:footnotedark}). This bound can be circumvented in the case of a pure millicharged DM without a dark photon, where the mDM annihilates only into SM photons and no extra degrees of freedom is introduced. For CDM, both BBN and $N_\text{eff}$ limits are model dependent; in the most conservative scenario, a mass of $m_\text{C}\gtrsim 400\text{ MeV}$ is sufficiently large for CDM to freeze out early enough to avoid any significant contribution to $N_{\text{eff}}$. To access lower masses, a less minimal model of CDM freeze-out should be explored.   

Throughout this paper we have neglected  heating of any of the three fluids from the annihilation of dark sector particles. Although previous work has shown that the annihilation of millicharged DM can have a non-negligible heating effect on the baryon temperature~\cite{DAmico:2018sxd,Liu:2018uzy,Mitridate:2018iag}, the annihilation of mDM is suppressed by $f_\text{m}^2$, and the value of $f_\text{m}$ examined here is small enough that no appreciable heating of the baryons is anticipated. Dark sector annihilations into other dark sector particles are also unlikely to have any effect on the temperature of the dark sector fluids. We refer the reader  to Appendix~\ref{sub:constraints} for more details on this topic.

\subsection{Neutral Hydrogen/Helium Scattering}
\label{sec:atoms}

With the range of viable mDM masses extended to 200 GeV, interactions between mDM and neutral hydrogen and helium become important to consider. Having an mDM mass heavier than the proton mass leads to a higher momentum transfer in the mDM-gas collision compared to the standard mDM setup where the bound in Eq.~\eqref{eq:upperboundmDM} applies. This enhances the mDM scattering with hydrogen and helium when the momentum transfer is larger than the inverse Bohr radius, resulting in an unscreened interaction with the nuclei of these atoms at energies much smaller than their ionization energies. In fact, the momentum transfer to helium atoms is comparable to that of hydrogen atoms despite the smaller abundance of helium, due to the larger nuclear charge and more diffuse electron wavefunction. The mDM scattering rates with atoms have a non-trivial velocity dependence, exhibiting a different behavior from the interactions of mDM with the ionized fraction. Including these interactions in our numerical calculation leads to changes of order ${\cal O}(5)$ in the value of $Q$ required to fit the EDGES result for mDM masses above \SI{100}{\mega\eV}. The relative importance of scattering with neutral (solid green line) vs. charged particles (solid orange line) for one particular choice of parameters is shown in the right panel of Fig.~\ref{fig:temp_ev}.

For the mDM charges considered here, the Born approximation breaks down at small enough velocities, and we expect resonance effects to enhance the low-redshift scattering of mDM with the atoms. The parametrics of when we expect this to happen is discussed in Appendix~\ref{sun:Beyond_Born}, and corresponds to the dotted-gray line in Fig.~\ref{fig:temp_ev}. At correspondingly small redshifts, the naive estimation of the atomic-mDM interaction rate should not be trusted, and large corrections may exist. Since this effect is unlikely to change our results qualitatively, and will only enhance the mDM-baryon scattering rate, thereby enlarging the viable parameter space further, we defer a more complete computation of this effect to future work. For a detailed discussion of how mDM-neutral atom scattering is included in our calculation, see Appendix~\ref{sec:study_case}.

\section{Phenomenology}
\label{sec:pheno}

The phenomenological consequences of our setup are very distinctive, and a large portion of the available parameter region will be probed in the near future. Beam dump and collider experiments can probe the region of parameter space with larger values of $Q$ by directly producing mDM, while direct detection experiments are sensitive to lower values of $Q$ for both mDM and CDM. The latter acquires a non-zero scattering cross section with the SM via a loop of mDM particles.

\subsection{Millicharged Dark Matter Component}
\label{sec:millipheno}

The constraints on the mDM component are summarized in Fig.~\ref{fig:money_milli}, with the viable parameter space shown in white.  Within that region the mDM charge, $Q$, is too small to be already excluded by present collider searches \cite{Davidson:2000hf,Badertscher:2006fm,Prinz:1998ua,CMS:2012xi} and too large for direct detection experiments to be sensitive. The green region in Fig.~\ref{fig:money_milli} shows the most recent determination of direct detection experimental sensitivity to particles with large $Q$, based on Ref.~\cite{Emken:2019tni}. The sensitivity of direct detection constraints in the large $Q$ region for mDM masses above 10 MeV goes down to fraction as small as $f_\text{m}=10^{-9}$ and therefore applies to the full parameter space considered here. Due to energy losses of mDM particles in the Earth overburden, the maximal reach for high-$Q$ is achieved by surface runs of the various experiments. For charges above that, the energy of most mDM particles reaching the Earth surface is below the energy threshold of single-electron scattering experiments due to atmospheric attenuation. At masses above 1 GeV, we extrapolate the determination of Ref.~\cite{Emken:2019tni} using their approximate analytic treatment of the energy losses\footnote{Since the mDM interactions with the SM are long-range, the elastic scattering is forward enhanced in the mass range we are considering, and the dominant shielding effect comes from the energy losses and not from reflection.}, which happens to reproduce the high-mass Monte Carlo results accurately.  

Present collider and beam dump experiments can probe the existence of millicharged particles regardless of their cosmological abundance. CMS can set constraints on particles with charges above $Q = 1/3$ by looking for soft tracks created by their interactions with the tracking material \cite{CMS:2012xi}. However, at lower $Q$, mDM pairs will depart the detector without interacting. The only way to constrain them would be to look at the missing energy given by their recoil against a hard object (either a photon or a jet) from initial state radiation. At the LHC, the reach of these search strategies is limited because of large backgrounds (see for instance the discussion in Ref.~\cite{Haas:2014dda}) and improvements are only possible at future lepton colliders. 

More interestingly, a robust search program for millicharged particles has been developed at beam dump experiments, following up on the first dedicated experiment on millicharged particles performed at SLAC~\cite{Prinz:1998ua} . The potential reach of the various proposed searches are summarized in Fig.~\ref{fig:money_milli}. These can be broadly divided into three classes: ionization-based experiments~\cite{Prinz:1998ua,Ball:2016zrp,Kelly:2018brz,Harnik:2019zee}, electron scattering experiments~\cite{Magill:2018tbb} and missing momentum experiments~\cite{Berlin:2018bsc}. The idea of most of these setups is to produce millicharged particles in SM particle collisions that are then detected downstream in scintillators~\cite{Prinz:1998ua,Ball:2016zrp}, liquid argon detectors~\cite{Harnik:2019zee}, or by exploiting their scattering with electrons~\cite{Magill:2018tbb}. Alternatively, missing momentum experiments such as LDMX~\cite{Berlin:2018bsc} can be sensitive to the emission of mDM pairs by measuring the time evolution of the momenta of electrons and muons as they pass through the detector.  On short timescales, the ArgoNeut collaboration~\cite{Anderson:2012vc} can almost completely exclude the EDGES explanation for an mDM fraction of $f_\text{m}\lesssim10^{-8}$, assuming that the result in Ref.~\cite{Harnik:2019zee} is confirmed. In the near future, the installation of milliQan at CERN~\cite{Yoo:2018lhk} can cover a large portion of the EDGES explanation up to a fraction $f_\text{m}\lesssim 10^{-6}$. Larger mDM fractions (and therefore smaller electric charges) can also be probed in the future with the installation of the Fermini detector on the NuMI beamline~\cite{Kelly:2018brz}, or in more ambitious future experimental facilities like DUNE and SHiP~\cite{Magill:2018tbb,Harnik:2019zee}.

As seen in Fig.~\ref{fig:money_milli}, the beam dump program has a limited reach for mDM charges smaller than $\sim10^{-3}$ and for mDM masses above 10 GeV. The direct detection program can in principle complement the collider effort in probing the mDM explanation of the EDGES anomaly completely. The main hurdle to overcome in this case is the mDM energy losses in the atmosphere. The atmosphere shielding can in principle be reduced by installing single-electron-threshold detectors on a balloon or on a satellite. In Fig.~\ref{fig:money_milli}, we show the projected sensitivity for a SENSEI-like dark matter detector installed on a balloon/satellite, with a 0.1 gram-month exposure taken from Ref.~\cite{Emken:2019tni}. Constraints from previous rocket experiments like XQC~\cite{Mahdawi:2018euy} are not sensitive to the small mDM fractions required to fit EDGES and hence do not appear in the plot. We also do not include the exclusion from RRS~\cite{Rich:1987st} which has not been properly recast for mDM. Alternatively, the experimental threshold of direct detection experiments can be reduced such that they become sensitive to the very slow mDM particles reaching the surface of the Earth. Given the existing proposals for low-threshold experiments below single-electron recoil energy \cite{Essig:2016crl,Schutz:2016tid,Knapen:2016cue,Budnik:2017sbu,Hochberg:2017wce,Bunting:2017net,Knapen:2017ekk,Rajendran:2017ynw,Griffin:2018bjn,Pospelov:2019vuf,Essig:2019kfe} and the flourishing ongoing activity in this direction, it would be interesting to study the ability of these experiments to probe the strongly interacting window for mDM. 

Further constraints on the mDM parameter space can come from astrophysical probes. Two examples in this direction are the constraints on mDM-baryon scattering which can be deduced from the heating of very cold gas clouds in the interstellar medium \cite{Bhoonah:2018wmw,Bhoonah:2018gjb} or in dwarf galaxies \cite{Wadekar:2019xnf}. These types of probes are  affected by large systematic uncertainties, but have in principle the sensitivity to probe a large portion of the mDM window displayed in Fig.~\ref{fig:money_milli}. Moreover, it is worth mentioning that the mDM component will also have non-trivial interactions with the magnetic fields in the interstellar medium  and could also be accelerated in supernova shocks like regular cosmic rays~\cite{Chuzhoy:2008zy}. A quantitative assessment of these effects, their astrophysical systematics and their dependence on the details of the dark sector is then necessary to fully understand the direct detection prospects in the parameter space shown in Fig.~\ref{fig:money_milli} \cite{toappearJ}.

Finally, another interesting problem would be to determine further consequences of our setup at lower redshifts. An example would be to understand to what extent the CDM distribution in the Milky Way could be modified by the presence of a small fraction of mDM strongly interacting with the baryons. Indeed, in the absence of mDM-CDM interactions, the mDM would thermalize with the baryons and be dragged into the Milky Way as first noticed in Ref.~\cite{DeRujula:1989fe}. Adding mDM-CDM interactions can substantially alter the standard mDM picture, allowing mDM to be coupled back to the CDM halo. A quantitative assessment of the consequences of our framework for structure formation goes beyond the scope of this work and is left for future studies.

\begin{figure*}
\centering
\includegraphics[width=0.48\textwidth]{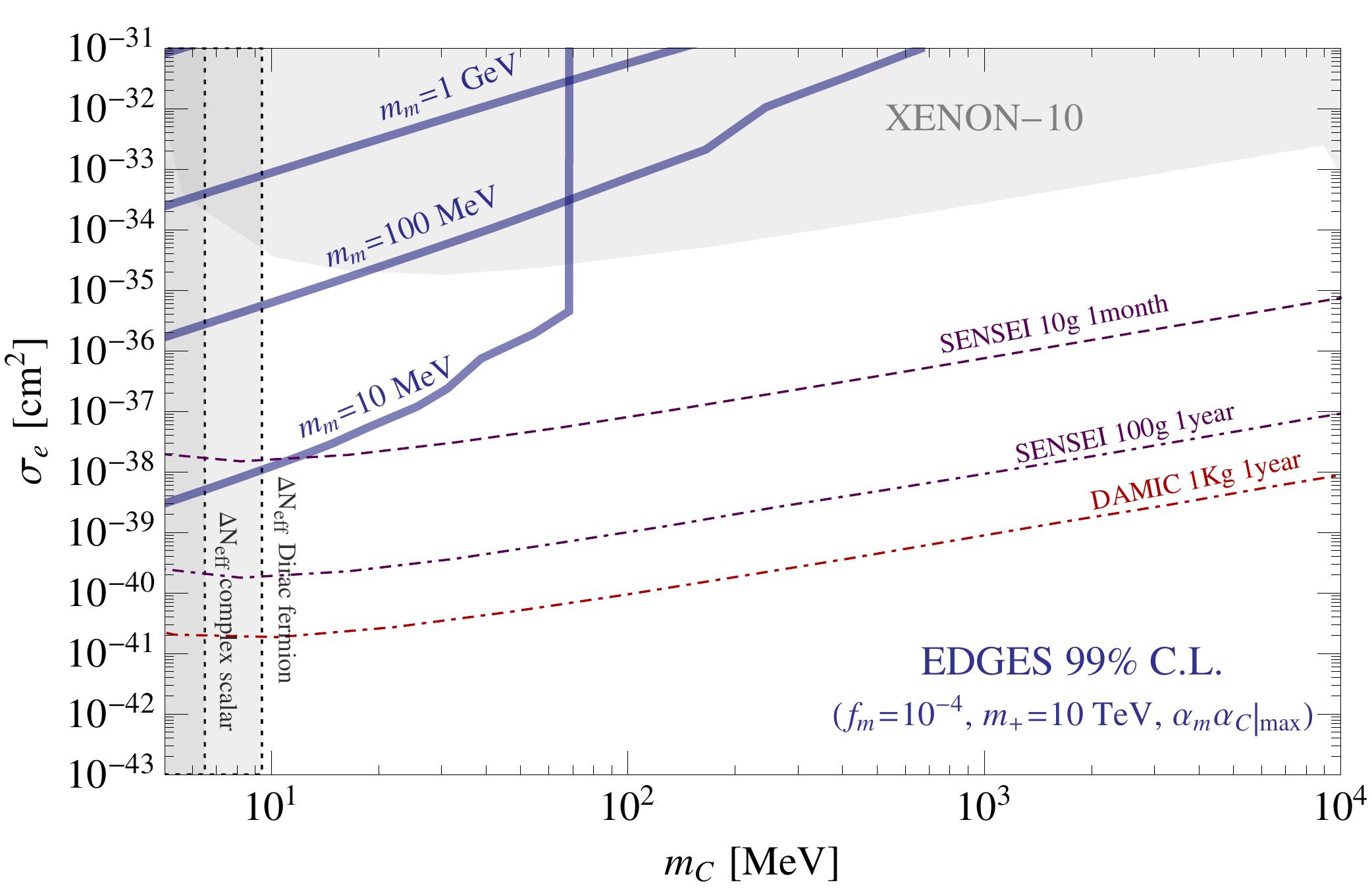}\;\;\includegraphics[width=0.48\textwidth]{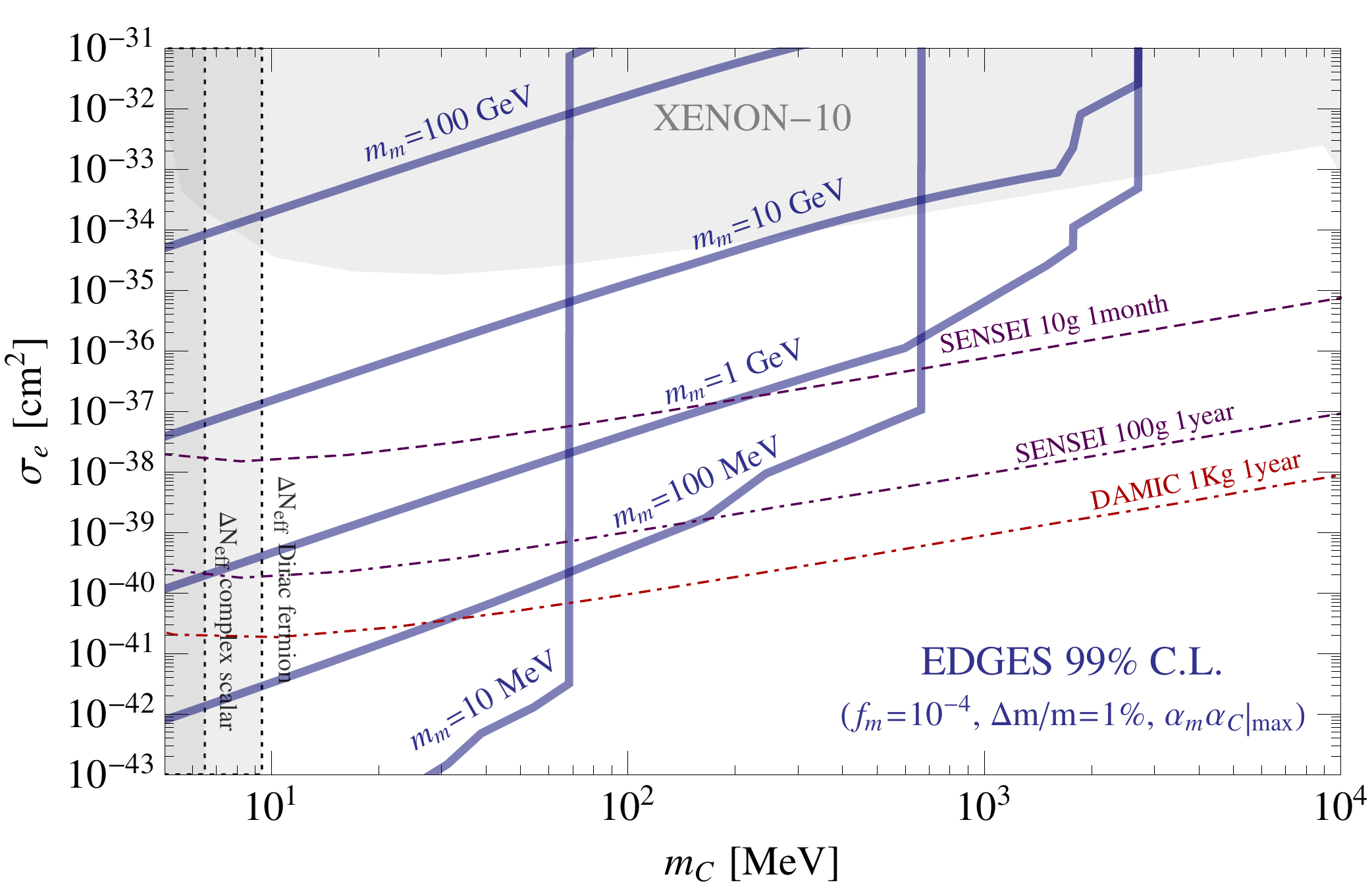}
\caption{Prospects for the direct detection of the dominant CDM component that couples to mDM via a light vector and acquires a coupling to the baryons through a loop of mDM. Gray shaded regions show the current bounds from Xenon-10 recast for electron scattering~\cite{Essig:2012yx} and the nuclear recoil bounds taken from Ref.~\cite{Essig:2015cda}. The vertical dotted lines at low masses indicate the $\Delta N_{\text{eff}}$ constraints from BBN for a Dirac fermion and a complex scalar extracted from \cite{Essig:2015cda}. All the sensitivities in the plot correspond to three signal events under the assumption of zero background. The dashed (dashed-dotted) purple line is the reach of SENSEI 10 g with 1 month exposure (100 g with 1 year exposure). The red dashed-dotted line gives the reach of DAMIC with 1 kg and 1 year of exposure. The thick blue lines are predictions for the direct detection cross section of the CDM component fitting EDGES for different mDM masses at fixed mDM fraction $f_\text{m}=10^{-4}$ and fixed $g_\text{m} g_\text{C}$ to be the maximal coupling satisfying the CMB constraint in Eq.~\eqref{eq:cmbconstrain} (see App.~\ref{sub:constraints} for further details). On the {\bf left} figure we assume a minimal model with a single light millicharged particle, where in Eq.~\eqref{eq:xsec} we fix $m_{+}=10\text{ TeV}$ and $m_{-}=m_{\text{m}}$ (see Appendix \ref{sec:bathmodels} for details). On the {\bf right} we show the case of a Dirac pair of millicharged particles with a mass splitting $\Delta m_{\text{m}}/m_{\text{m}}=10^{-2}$.  Correspondingly, $\bar{\sigma}_e$ is suppressed by $\left(\Delta m/m\right)^2=10^{-4}$  with respect to the naive log-enhanced expectation (see Appendix \ref{sec:bathmodels} for details).}   
\label{fig:money_CDM_vector}  
\end{figure*}

\subsection{Cold Dark Matter Bath}
\label{sec:CDMpheno}

In our setup, the CDM bath interacts with mDM via a long-range interaction. This implies the presence of long-range CDM self-interactions mediated by the same force carrier. Accounting for the most conservative constraints on CDM self-interaction from Refs.~\cite{Buckley:2009in,Agrawal:2016quu}, we require
\begin{equation}
	\frac{\alpha_{\text{C}}^2}{m_{\text{C}}^3}\lesssim 10^{-11}\,\rm GeV^{-3}.
\end{equation} 
We emphasize, however, that the robustness of such constraints is still under debate (see e.g.\ Ref.~\cite{Peter:2012jh}). 

CDM-SM interactions are radiatively generated through a loop of mDM pairs. For a given choice of mDM mass, mDM fraction, and mDM-SM and mDM-CDM couplings (possibly to explain the EDGES result) one expects a minimal CDM-SM cross section. To illustrate this point, we adopt a simple benchmark model where a
light vector boson mediates the mDM-CDM interactions. The direct detection cross section of CDM is induced by the mixing between the new vector boson and the SM photon generated by loops of mDM particles, and is proportional to $Q^2\alpha_\text{C} \alpha_\text{m} $.
For a choice of mDM mass, $m_{\rm m}$, and fraction $f_{\rm m}$, and  fixing the value of $g_\text{m} g_\text{C}$ to saturate the CMB constraint, the EDGES signal fixes $Q$, as shown in Fig.~\ref{fig:money_milli}. Using this value, we get a prediction for the direct detection cross section of CDM. 
For an mDM sector containing two Dirac fermions with masses $\sim m_{\pm}$, the direct detection cross section defined in Eq.~\eqref{eq:defsigmae} reads
\begin{equation}
\bar{\sigma}_e\simeq
\frac{8 Q^2g_\text{C}^2 g_\text{m}^2}{g_{\text{EM}}^4 \mu_{e\text{m}}^2}\left(\log\frac{m_+}{m_-}\right)^2 , \label{eq:xsec}
\end{equation}
where we neglect the log-enhancement of the Rutherford scattering. When $m_+\gg m_-$ we recover, the UV dependence of the cross section in the case of a single Dirac pair (see Appendix~\ref{sec:bathmodels}). If instead we define $\Delta m_\text{m}=(m_+-m_{-})/2$ and $m_\text{m}=(m_++m_{-})$, we see that $\log(m_+/m_-) \sim \Delta m_\text{m} /m_\text{m}$ in the limit $\Delta m_\text{m} \ll m_\text{m}$ and the direct detection cross section is then suppressed as $(\Delta m_\text{m}/ m_\text{m})^2$ for small splittings between the mDM Dirac pairs.  The detailed derivation of the CDM cross section and its dependence on the structure of the mDM sector as well as on $g_\text{m} g_\text{C}$ is discussed in Appendix~\ref{sec:bathmodels}. 

In Fig.~\ref{fig:money_CDM_vector} we show the CDM direct detection cross section for  $f_{\rm m}=10^{-4}$ with the maximal  $g_\text{m} g_\text{C}$ allowed by CMB and different choices of the mDM mass, with $Q$ chosen as a function of $m_\text{m}$ and $m_\text{C}$ to account for the EDGES signal.  On the left panel, we consider the simplest model, where the UV cut-off is fixed at $m_{+}=10 \text{ TeV}$ and $m_{-}\equiv m_{\text{m}}$. In this scenario, for $m_\text{m}$ masses larger than  100 MeV, the direct detection cross section of CDM is already excluded by Xenon-10 data~\cite{Essig:2012yx}. For mDM masses below 100 MeV, the direct detection of CDM cross section is within the reach of the future SENSEI with10~gram of detector material and 1 month of exposure. On the right panel, we show a scenario where $\Delta m_\text{m}/m_\text{m}\simeq 10^{-2}$, which is of the order of the splitting between charged and neutral pions. In this case, the direct detection cross section is suppressed by four orders of magnitude; all of the mDM masses below 10 GeV give a currently viable cross section that is within the reach of future low-threshold DM experiments.

\section{Conclusion}
\label{sec:conclusions}

We present a new setup where millicharged DM-baryon scattering produces large effects that can be observed in 21-cm cosmology, without being constrained by any current cosmological, terrestrial or astrophysical probe. This work has been motivated by the recent EDGES measurement~\cite{Bowman2018}, which exhibits an anomalously large absorption peak in the 21-cm global spectrum at the cosmic dawn. In our model, the cooling capability of the millicharged DM (mDM) is enhanced by introducing a long-range interaction between the cold dark matter (CDM) component and the mDM fraction, thereby allowing the CDM component to act as a heat sink while only interacting weakly with the SM. As a consequence, and in contrast  to previous attempts at explaining the EDGES result with mDM, the mDM parameter space extends to mDM masses as high as  200 GeV, with a dark matter energy density fraction as small as $10^{-8}$, thereby relaxing the existing bounds considerably. 

The viable parameter space for which a fraction of mDM can explain EDGES in our setup can be effectively probed by the rich collider experimental program aimed at searching for millicharged particles at beam dump facilities~\cite{Ball:2016zrp,Magill:2018tbb,Kelly:2018brz,Berlin:2018bsc,Harnik:2019zee}. Moreover, future single-electron-threshold direct detection experiments above ground~\cite{Emken:2019tni} (either in balloon or satellites), together with ground-based direct detection experiments with energy threshold below the single electron recoil~\cite{Essig:2016crl,Schutz:2016tid,Knapen:2016cue,Budnik:2017sbu,Hochberg:2017wce,Bunting:2017net,Knapen:2017ekk,Rajendran:2017ynw,Griffin:2018bjn,Pospelov:2019vuf,Essig:2019kfe} serve as a complementary probe at lower mDM charges.

Interestingly, fitting EDGES in our setup provides a surprising connection between a low-redshift cosmological signal and the direct detection of the CDM bath. We have shown how forthcoming direct detection experiments sensitive to electron recoils, such as SENSEI and DAMIC~\cite{Battaglieri:2017aum}, will be able to probe a large portion of the CDM bath parameter space which explains the EDGES result. In particular, the viable EDGES parameter space for the simplest realization of our setup should be probed in the near future by SENSEI-10g and SENSEI-100g. 

Beyond providing a potential explanation for the EDGES result,  we have shown that the inclusion of the scattering of mDM with both hydrogen and helium in the Born approximation dominates the cooling rate for sufficiently heavy mDM, and should be included to correctly compute the 21-cm signal~\cite{Munoz:2015bca}. Extending our treatment beyond the Born approximation and accounting for the effects of mDM-SM resonances that are expected to appear goes beyond the scope of this paper and is postponed to future work. Including this effect together with a more complete modelling of astrophysical effects at low redshifts~\cite{Madau:1996cs,Chen:2003gc,Cohen:2016jbh,Venumadhav:2018uwn} would be necessary to better assess the predictions of our setup for the absorption peak shape at cosmic dawn. We hope to come back to this issue in the near future.  

\section*{Acknowledgements}
We thank Tejaswi Venumadhav for collaborating during the early stages of this project, Josh Ruderman and Erik Kuflik for many useful comments on this work and Rouven Essig and Timon Emken for correspondence on Ref.~\cite{Emken:2019tni}. We also thank Raffaele Tito D'Agnolo, Marat Freytsis, Simon Knapen, Gordan Krnjaic, Zhen Liu and Harikrishnan Ramani for interesting thoughts on the many possible experimental probes of our setup. 
DR and TV  thank MIAPP for generous hospitality during the completion of this work. NJO, DR and TV further thank the generous hospitality of Neot Smadar during the completion of this work. HL is supported by funds from the Office of High Energy Physics of the U.S. Department of Energy under Grant No. DE-SC00012567 and DE-SC0013999.  NJO is grateful to the Azrieli Foundation for the award of an Azrieli Fellowship.
TV is  supported in part by the Israel Science Foundation-NSFC (grant No. 2522/17), by the European Research Council (ERC) under the EU Horizon 2020 Programme (ERC- CoG-2015 - Proposal n.~682676 LDMThExp), and by a grant from the Ambrose Monell Foundation, given by the Institute for Advanced Study.

\bibliography{21.bib}

\newpage
\onecolumngrid
\appendix

\section{Temperature Evolution Formalism} 
\label{sec:21cm}

In this appendix, we present a general treatment for the temperature evolution of two non-relativistic fluids interacting through elastic processes. As previously derived in Refs.~\cite{Tashiro:2014tsa,Munoz:2015bca}, the thermal evolution of the two sectors is sensitive to two macroscopic quantities: the heat transfer rate between the two fluids, $\dot Q$, and the drag between the two fluids due to relative bulk motion, quantified by the "drag term" $D$ which is generally a function of $V_{\rm rel}$, the relative velocity between the fluids. The system is described by three thermodynamic parameters: the temperature of each fluid and the relative bulk motion between the two fluids. 
Denoting the two sectors by subscripts 1 and 2, the thermodynamics of the two fluids is determined by the following set of coupled non-linear equations:
\begin{align} \label{eq:T_eom}
\frac{d T_1}{d \log a} &=-2 T_1+\frac{2}{3} \frac{\dot{Q}_1}{H}\,,\\ 
\frac{d T_2}{d \log a} &=-2 T_{2}+\frac{2}{3} \frac{\dot{Q}_{2}}{H}\,,\\ 
\frac{d V_{\rm { rel }}}{d \log a} &=-V_{\rm { rel }}-\frac{D}{H}\,.
\end{align}
Here $a$ is the scale factor and $H$ is the Hubble parameter.  The non-linearities arise because  $\dot Q$ and $D$ are non-linear functions of the temperatures and the velocity. To solve these equations one needs to specify $\dot Q$ and $D$, which are related to the  microscopic properties of the interaction between the particles in the two gases. This is the subject of the next subsection. 

\subsection{Heat Transfer Rate and Drag Term} 
\label{sub:calculating_thermodynamic_quantities}

We are interested in calculating two macroscopic quantities, the heat transfer rate and the drag term. We relate those quantities to the underlying particle physics through the thermal average of energy and momentum that is transferred when particles in each fluid scatter off each other elastically. We will calculate everything in the rest frame of fluid 1. Galilean invariance allows the relevant quantities for fluid 2 to be obtained by simply substituting $1\leftrightarrow 2$.
The thermally averaged momentum and energy transfer from fluid 2 to fluid 1 is given by
\begin{equation}
    \dot {\vec{P}}_{1}=\left\langle\int d \Omega \, n_{2} \frac{d \sigma}{d \Omega} \vec{q} v_{\mathrm{rel}}\right\rangle \quad , \quad \dot{Q}_{1}=\left\langle\int d \Omega \, n_{2} \frac{d \sigma}{d \Omega} \Delta E_{1} v_{\mathrm{rel}}\right\rangle\,,
\end{equation}
where $\vec {q}$ is the momentum transfer in the collision, $\Delta E_1=\vec{v}_{\rm CM}\cdot \vec{q}$, $\vec{v}_{\rm CM}$ is the center-of-mass velocity of the two particles in each interaction, and $n_2$ is the number density in fluid 2. The symbol $\langle \cdot\rangle$ is the thermal average. The momentum transfer is related to the scattering angle through $ {q}^2=2 \mu^2v_{\rm rel}^2(1-\cos \theta)$, with $\mu$ the reduced mass and $v_{\rm rel}$ the relative velocity between the colliding particles.  
The transfer cross section is defined by~\cite{pitaevskii2012physical}
\begin{equation}
     \sigma_{T}\vec{p}_{\rm rel}=\int d \Omega \frac{d \sigma}{d \Omega} \vec{q}\;\;,\;\;\vec{p}_{\rm rel}=\mu \vec{v}_{\rm rel} \,.\label{eq:deftransf}
\end{equation}
Using the kinematics relation $2\vec{p}_{\rm rel} \cdot \vec{q}=-q^2$, we can write 
\begin{equation}\label{eq:sig_T}
    \sigma_T(p_{\rm rel})=\frac{\pi}{2 p_{\rm rel}^{4}} \int_{0}^{4 p_{\rm rel}^{2}} d q^{2} \, q^{2} \frac{d \sigma}{d \Omega}.
\end{equation}
With this notation, both the averaged momentum transfer and energy transfer can be expressed simply as
\begin{equation}
    \dot {\vec{P}}_1=  \left\langle \Gamma_1\vec{p}_{\rm rel}\right\rangle\;\;,\;\;\dot{Q}_1=  \left\langle \Gamma_1\vec{v}_{\rm CM}\cdot\vec{p}_{\rm rel}\right\rangle,
\end{equation}
with $\Gamma_1=n_2 \sigma_T(v_{\rm rel})v_{\rm rel}$ being the rate of collisions of particle 1 with particle 2 and $\sigma_T(v_{\rm rel})$ is defined by Eq.~\eqref{eq:sig_T} substituting $\vec{p}_{\text{rel}}=\mu \vec{v}_{\text{rel}}$. 

Momentum conservation ensures that $n_1\dot {\vec{P}}_1+n_2\dot {\vec{P}}_2=0$; we use this to calculate the rate of change in the relative  bulk motion,
\begin{equation}
    \dot{\vec{V}}_{\rm rel}=\frac{\dot {\vec{P}}_1}{m_1}-\frac{\dot {\vec{P}}_2}{m_2}=\frac{\dot {\vec{P}}_1}{m_1}\left(1+\frac{\rho_1}{\rho_2}\right)=\frac{\rho_1+\rho_2}{m_1+m_2}\langle \sigma_T v_{\rm rel}\vec{v}_{\rm rel}\rangle \,.
\end{equation}
It is convenient to use a scalar quantity to quantify the rate of change of relative velocity, defined as
\begin{equation}\label{eq:drag}
    D\equiv \dot{V}_{\rm rel}=\frac{\vec{ V}_{\rm rel}\cdot \dot{\vec{V}}_{\rm rel}}{{ V}_{\rm rel}}=\frac{1}{{ V}_{\rm rel}}\frac{\rho_1+\rho_2}{m_1+m_2}\left\langle \sigma_T v_{\rm rel}\vec{v}_{\rm rel}\cdot\vec{V}_{\rm rel}\right\rangle\,.
\end{equation}

To simplify the heat transfer rate we decompose $\vec{v}_{\rm CM}$ into three directions: $\vec{V}_{\rm rel}$, $v_-\equiv \vec{v}_{\rm rel}-\vec{V}_{\rm rel}$ and $\vec{v}_{\rm rand}$, with the latter accounting for thermal fluctuations. The center-of-mass velocity is simply
\begin{equation}
    \vec{v}_{\rm CM}=\frac{m_1 \vec{v}_1+m_2 \vec{v}_2}{m_1+m_2} \,,
\end{equation}
where we denote the velocities of particles 1 and 2 by $\vec{v}_{1,2}$.
Since the relative velocity is given by $\vec{v}_{\rm rel}=\vec{v}_1-\vec{v}_2$, the random component (in the rest frame of fluid 1) must be
\begin{equation}
    \vec{v}_{\rm { rand }}=\frac{\frac{m_{1}}{T_{1}} \vec{v}_{1}+\frac{m_{2}}{T_{2}}\left(\vec{v}_{2}-\vec{V}_{\rm { rel }}\right)}{\frac{m_{1}}{T_{1}}+\frac{m_{2}}{T_{2}}} \,.
\end{equation}
In this new basis the centre of mass velocity is given by
\begin{equation}
    \vec{v}_{\rm CM}=\vec{v}_{\rm { rand }}+\frac{\mu}{m_{1}} \vec{V}_{\rm { rel }}+\frac{T_{2}-T_{1}}{\left(m_{1}+m_{2}\right) u_{\rm { th }}^{2}} \vec{v}_{-},
\end{equation}
with $u_{\rm th}^2\equiv\left( \langle v_{\text{rel}}^2\rangle -\langle \vec{v}_{\text{rel}}\rangle^2\right)/3=T_1/m_1+T_2/m_2$ being the variance of the relative velocity distribution. In the absence of relative motion and temperature difference ($\vec{V}_{\rm { rel }}=0$ and $T_1=T_2$) the fluid temperatures only evolve under adiabatic expansion. In such a situation, $\vec{v}_{\rm CM}=\vec{v}_{\rm rand}$, and $\dot{Q}_1=\left\langle \Gamma_1\vec{v}_{\rm rand}\cdot\vec{p}_{\rm rel}\right\rangle$ has to vanish. We thus conclude that $\vec{v}_{\rm rand}$ will not contribute to the heat transfer rate, and therefore

\begin{equation}
    \dot{Q}_{1}=\frac{\rho_{2}}{\rho_{1}+\rho_{2}} \mu  D V_{\mathrm{rel}}+\frac{T_2-T_1}{m_1+m_2}\frac{\mu}{u_{\rm th}^2}\left\langle \Gamma_1 \vec{v}_{\rm rel} \cdot \left(\vec{v}_{\rm rel}-\vec{V}_{\rm rel}\right)\right\rangle.
\end{equation}
We note that both $D$ and $\dot{Q}_1$ are now determined as thermal averages over the relative velocity $\vec{v}_{\rm rel}$ only. The $\vec{v}_{\rm rel}$ distribution is given by
\begin{equation}
    f(\vec{v}_{\rm rel})=(2 \pi u_{\rm th}^2)^{-3/2}\exp\left[-\frac{\left(\vec{v}_{\rm rel}-\vec{V}_{\rm rel}\right)^2}{2u_{\rm th}^2}\right].
\end{equation}
This shows that our results, as promised, are Galilean invariant, allowing us to infer $\dot{Q}_2$ by symmetry arguments. Moreover, energy conservation requires 
\begin{equation}\label{eq:e_conserv}
    n_{1} \dot{Q}_{1}+n_{2} \dot{Q}_{2}=\frac{\rho_{1} \rho_{2}}{\rho_{1}+\rho_{2}} V_{\mathrm{rel}} D \,.
\end{equation}

\subsection{Form Factor Integrals} 
\label{sub:form_factors_integrals}

The thermodynamic functions $\dot {Q}$ and $D$ defined above are given in terms of the following two integrals:
\begin{align}
    &I_D\equiv\left\langle \sigma_T v_{\rm rel}\vec{v}_{\rm rel}\cdot\vec{V}_{\rm rel}\right\rangle=\int d^{3} v_{\mathrm{rel}} f(v_{\rm { rel }}) v_{\mathrm{rel}} \vec{v}_{\mathrm{rel}} \cdot \vec{V}_{\mathrm{rel}} \sigma_{T}\left(v_{\mathrm{rel}}\right) \,, \nonumber\\
    &I_T\equiv\left\langle \sigma_T v_{\rm rel}\vec{v}_{\rm rel}\cdot\left(\vec{v}_{\rm { rel }}-\vec{V}_{\rm { rel }}\right)\right\rangle=\int d^{3} v_{\rm { rel }} f(v_{\rm { rel }}) v_{\rm { rel }} \vec{v}_{\rm { rel }} \cdot\left(\vec{v}_{\rm { rel }}-\vec{V}_{\rm { rel }}\right) \sigma_{T}\left(v_{\rm { rel }}\right).
\end{align}
To present this in a more model-independent fashion, we introduce the cross section parametrization adopted by direct detection experiments~\cite{Essig:2011nj}:
\begin{equation}\label{eq:sig_to_ff}
    \frac{d \sigma}{d \Omega}=\frac{\bar{\sigma}}{4 \pi}\left|F\left(q^{2}\right)\right|^{2}=\frac{\bar{\sigma}}{4 \pi}\left|f_{\rm D M}\left(q^{2}\right)\right|^{2} \left|f_{\rm S M}\left(q^{2}\right)\right|^{2}\, ,
\end{equation}
where $\bar \sigma$ is a constant and $F(q^2)$ is the interaction form factor, further decomposed into DM and SM contributions, written as $f_\text{DM}(q^2)$ and $f_\text{SM}(q^2)$ respectively. Using this notation together with the definition of $\sigma_T$ in Eq.~\eqref{eq:sig_T}, and changing the order of integration so that we integrate first over $v_{\rm rel}$ and then over $q^2$, we arrive at the simple 1D integrals below:
\begin{align}
    &I_{T}=\frac{\bar{\sigma} u_{\mathrm{th}}}{\sqrt{8 \pi} V_{\mathrm{rel}} \mu^{3}} \int_{0}^{\infty} d q \, q^{2}\left|F\left(q^{2}\right)\right|^{2}\left\{\exp \left[-\left(\frac{2 \mu V_{\mathrm{rel}}-q}{2 \sqrt{2} \mu u_{\mathrm{th}}}\right)^{2}\right]-(q \rightarrow-q)\right\} \,, \label{eq:int1}\\
    &I_{D}=-I_T+\frac{\bar{\sigma}}{8 V_{\mathrm{rel}} \mu^{4}} \int_{0}^{\infty} d q \, q^{3}\left|F\left(q^{2}\right)\right|^{2}\left[\operatorname{erf}\left(\frac{q+2 \mu V_{\mathrm{rel}}}{2 \sqrt{2} \mu u_{\mathrm{th}}}\right)+(q \rightarrow-q)\right] \,.\label{eq:int2}
\end{align} 
This is as far as we develop the formalism on model independent grounds; to proceed further, one should specify a model and calculate the form factor $F$, as we will do in the next section. For completeness, we give the expressions for the integrals above in the limit where $V_{\rm rel}\to 0$ and $V_{\rm rel}\to \infty$ in which only thermal motion and bulk motion matters respectively:
\begin{align}
    \lim _{V_{\rm{rel} \rightarrow 0}} I_{T}=\frac{\bar{\sigma}}{2 \sqrt{8 \pi} u_{\mathrm{th}} \mu^{4}} \int_{0}^{\infty} d q^{2} \, q^{2}\left|F\left(q^{2}\right)\right|^{2} \exp \left[-\left(\frac{q^{2}}{8 \mu^{2} u_{\mathrm{th}}^{2}}\right)\right]\;\;&, \;\;\lim _{V_{\mathrm{rel} \rightarrow 0}} I_{D}=0\label{eq:V_to_0}\, ,\\
    \lim _{u_{\mathrm{th}} \rightarrow 0} I_{D}=\frac{\bar{\sigma}}{8 V_{\mathrm{rel}} \mu^{4}} \int_{0}^{4 \mu^{2} V_{\mathrm{rel}}^{2}} d q^{2} \, q^{2}\left|F\left(q^{2}\right)\right|^{2}=V_{\mathrm{rel}}^{3} \sigma_{T}\left(V_{\mathrm{rel}}\right)\;\;&, \;\;\lim _{u_{\mathrm{th}} \rightarrow 0} I_{T}=0\, . \label{eq:V_to_inf}
\end{align}

\section{Millicharged Dark Matter Form Factors} 
\label{sec:study_case}
We now provide the complete form factors needed to correctly compute the temperature evolution of both the baryons and the mDM after recombination. The baryonic bath after recombination is composed of 93\% hydrogen and 7\% helium by number density, with roughly $10^{-4}$ of hydrogen being ionized in the form of H$^+$ and $e^-$. The interactions of mDM with the SM bath will then depend on the behavior of the form factors of the different components. 
We begin by presenting the form factors derived using  the Born approximation and then discuss the range of validity for this approximation.

\subsection{Born Approximation} 
\label{sub:ionized_fraction}

\subsubsection{Ionized Fraction} 
The differential ${\rm mDM}-I$ low energy scattering cross section with $I=e,p$ is given by
\begin{equation}
    \frac{d \sigma_{I}}{d \Omega}=4Q^2\alpha_{\rm EM}^2\frac{\mu_I^2}{(m_\phi^2+q^2)^2} \,,
\end{equation}
where $\mu_I$ is the $I$-mDM reduced mass and $m_\phi$ is the mediator mass, which we always take to be much smaller than the typical momentum exchange $q$. Using the direct detection experiments notation introduced in Ref.~\cite{Essig:2011nj}, we have
\begin{equation}
\bar{\sigma}_e=\frac{16 \pi \mu_e^2 Q^2\alpha_{\rm EM}^2}{(\alpha_{\rm EM} m_e)^4} \,,
\end{equation}
where $\mu_e$ is the $e$-mDM reduced mass. This allows us to write
\begin{equation}
    \frac{d \sigma_{I}}{d \Omega}=\frac{\bar{\sigma}_e}{4 \pi}\frac{\mu_I^2}{\mu_e^2}\frac{(\alpha_{\rm EM} m_e)^4}{(m_\phi^2+q^2)^2},\label{eq:defsigmae}
\end{equation}
We can thus identify the DM form factor defined in Eq.~\eqref{eq:sig_to_ff} as
\begin{equation}
    f^I_{\rm DM}(q^2)=\frac{\mu_I}{\mu_e}\frac{(\alpha_{\rm EM} m_e)^2}{(m_\phi^2+q^2)}.
\end{equation}
while the SM form factor is equal to one since we are dealing with point particles. 
\subsubsection{Atoms} 
To calculate the mDM-atom interaction, we will treat each atom as a positively charged point particle surrounded by a negative charge density,
\begin{equation}\label{eq:H_charge_density}
    \rho=Z e\left(\delta^3(\vec{r})-|\psi_{A,Z}(\vec{r})|^2\right),
\end{equation}
where $\psi_{A,Z}$ is the electron wave function which we assume to be hydrogen-like, and $Z$, $A$ are the atomic number and mass respectively of the atom under consideration. Note that the total charge when integrated over all space vanishes, but the negative charge is distributed over the support of the wave function. Within the Born approximation, this charge density leads to a differential cross section which can be factorized into an interaction with a point particle of mass $A m_p$ and charge $Z$ times a form factor.  The latter is the Fourier transform of the squared density above, 
\begin{equation}
    \frac{d \sigma_{A,Z}}{d \Omega}=\frac{\bar{\sigma}_e}{4 \pi}\left|f^{A}_{\rm DM} \left(q^2\right)\right|^2\left|f_{\rm SM}^{A,Z} \left(q^2\right)\right|^2,
\end{equation}
with
\begin{equation}\label{eq:A_FF}
     f^A_{\rm DM}(q^2)=\frac{\mu_A}{\mu_e}\frac{(\alpha_{\rm EM} m_e)^2}{(m_\phi^2+q^2)}\;\;,\;\;f_{\rm SM}^{A,Z}(q^2)=Z\left\{1-\left[1+\left(a_{\rm eff}\frac{q}{2}\right)^2\right]^{-2}\right\} \,.
\end{equation}
Here $a_{\rm eff}$ is the effective Bohr radius, i.e.\ $a_0 \equiv (\alpha_{\rm EM}m_e)^{-1}$ for hydrogen and roughly $a_0/1.69$ for helium~\cite{jackiw1997intermediate}. Since the mediators we consider are very light and will always obey $a_0 m_\phi \ll 1$, we can safely set $m_\phi = 0$ for all mDM-atom interactions.

\subsection{Beyond the Born Approximation}
\label{sun:Beyond_Born}
\subsubsection{Ionized Fraction} 
For scattering in a Yukawa potential, at low momentum transfer compared to the  Yukawa potential energy, the Born approximation no longer faithfully describes the scattering process~\cite{Buckley:2009in,Tulin:2013teo}. As was further shown in Ref.~\cite{Tulin:2013teo}, when the momentum transfer is also much larger than the Yukawa potential length-scale, the full non-perturbative quantum mechanical calculation reduces to the classical scattering calculation of Refs.~\cite{Khrapak:2003kjw,1308514}. The authors of Refs.~\cite{Khrapak:2003kjw,1308514} calculated the center-of-mass scattering process for an attractive Yukawa potential 
\begin{equation}
    V(r)=-\frac{ Q \alpha_\text{EM}}{r}e^{-m_\phi r}
\end{equation}
and found that for $Q \alpha_\text{EM}m_\phi\ll \mu v^2$, the transfer cross section defined in Eq.~\eqref{eq:deftransf} is very precisely approximated by
\begin{equation}
    \sigma^{\rm class}_T=2\cdot\frac{2 \pi  Q^2 \alpha_\text{EM}^2}{\mu^2 v^4}\log\left(\frac{\mu v^2}{Q\alpha_\text{EM}m_\phi}\right).
\end{equation}
We have reproduced this calculation also for a repulsive potential and obtained the same result in the same limit.
Comparing this with the Born approximation transfer cross section
\begin{equation}
    \sigma^{\rm Born}_T=\frac{2 \pi  Q^2 \alpha_\text{EM}^2}{\mu^2 v^4}\log\left(\frac{4\mu^2 v^2}{em_\phi^2}\right),
\end{equation}
we see that the classical \textit{differential} cross section can be obtained from the Born approximation through
\begin{equation}
    \frac{d \sigma_{\rm class}}{d \Omega}=2\cdot 4 Q^2 \alpha_\text{EM}^2\frac{\mu^{2}}{\left(m_{\rm eff}^{2}+q^{2}\right)^{2}}\;\;,\;\;m_{\rm eff}^2\equiv \frac{4}{e}Q \alpha_\text{EM} m_\phi \mu\,,
\end{equation}
where $e \simeq 2.71828$ is Euler's number. From this result we can extract the classical form factor in Eqs.~(\ref{eq:int1}) --~(\ref{eq:int2}):
\begin{equation}
    f^I_{\rm DM,\rm class}(q^2)=2\frac{\mu_I}{\mu_e}\frac{(\alpha_{\rm EM} m_e)^2}{(m_{\rm eff}^2+q^2)}\,.
\end{equation}

\subsubsection{Atoms} 
The scattering of mDM with hydrogen-like atoms requires extra care. The classical central potential between an mDM particle and the Hydrogen-like charge density of Eq.~\eqref{eq:H_charge_density} is given by
\begin{equation}
    V(r)=\pm QZ\alpha_{\rm EM}\frac{e^{-2r/a_{\rm eff}}}{r}\left(1+\frac{r}{a_{\rm eff}}\right),\label{eq:potential}
\end{equation}
where the $\pm$ sign corresponds to a repulsive and an attractive interaction respectively, and $a_{\rm eff}$ is the effective radius defined below Eq.~\eqref{eq:A_FF}. When a particle is close enough to feel the potential, the above potential is of order $Q \alpha_{\rm EM}/a_{\text{eff}}$. If this potential energy is larger than the kinetic energy in the scattering process we expect resonances to appear; this happens whenever
\begin{equation}
    \frac{Q}{v^2_{\text{rel}}}\gtrsim \frac{\mu_{\rm m}}{Z\alpha_{\rm EM}^2m_e}\sim 10^7\ ,\label{eq:boundstates}
\end{equation}
where $\mu_m$ is the reduced  mDM-atom mass. While the above inequality identifies when resonances are expected in the spectrum,
by computing when the second term in the Born series becomes as important as the first, we get a self-consistency condition for the Born approximation to hold~\cite{nla.cat-vn1011842}
\begin{equation}
    \frac{\mu_{\rm m} }{2 \pi }\left| \int d^3 r e^{i {\mu_{\rm m} v_{\rm rel} r} \left(c_{\theta }+1\right)}\frac{ V(r) }{r}\right| \ll1\, .
\end{equation}
This can be calculated analytically for the potential in Eq.~\eqref{eq:potential} and in the limit $\mu_{\rm m} v_{\rm m} a_{\text{eff}}\gg1$ reduces to
\begin{equation}
    \frac{Z Q \alpha_{\rm EM}}{v_{\text{rel}}}\log\left(\mu_\text{m} v_{\text{rel}} a_{\text{eff}}\right)\ll 1\,.
\end{equation}
During the cosmic dawn, the relative velocity between the mDM and the baryonic bath can be as low as $v_{\text{rel}} \sim 10^{-6}$. As a consequence the Born approximation breaks down for a large portion of the allowed parameter space in Fig.~\ref{fig:money_milli}. A typical example of when the Born approximation breaks down is indicated in Fig.~\ref{fig:temp_ev}. Including a full quantum computation of the scattering mDM-atoms would enhance the cooling rate at low redshift, further enlarging the parameter space presented in our study. We defer a detailed study of this effect for a future publication.

\section{Details of the Framework} 
\label{sec:analytical}
We now present the precise equations at the heart of this paper governing the evolution of all the relevant physical quantities in our model. 
We start by generalizing the thermal evolution equations presented in App.~\ref{sec:21cm} to include three different fluids and study the different regimes. We discuss in particular how our new solution is parametrically connected to the standard mDM scenario.  
Lastly, we discuss several important constraints on these parameters, including a simple way of recasting CMB constraints for our model.   

\subsection{Three Fluid System and Parameter Space} 
\label{sub:three_fluids_system}

We are interested in the thermodynamic evolution equation of three coupled fluids: CDM, mDM and baryons. For simplicity we will assume throughout that the different constituents of the baryonic fluid always maintain thermal equilibrium. The thermodynamic variables in that case are the temperature of each fluid $T_{\rm C},\,T_{\rm m}$ and $T_{\rm b}$, and two relative bulk velocities $\vec{V}_{\rm bm}$ and $\vec{V}_{\rm mC}$. We assume for simplicity that the two relative velocities lie along the same direction, which is justified by our boundary conditions, which assume mDM to be tightly coupled to the baryonic fluid at high redshift. The thermodynamic evolution is then given by
\begin{align} 
&\frac{d T_{\rm b}}{d \log a} =-2 T_{\rm b}+\frac{2}{3} \frac{\dot{Q}_{\rm bm}}{H}+\frac{\Gamma _{\rm Comp}}{H}\left(T_\gamma-T_{\rm b}\right),\\ 
&\frac{d T_{\rm m}}{d \log a} =-2 T_{\rm m}+\frac{2}{3} \frac{\dot{Q}_{\rm mC}}{H}+\frac{2}{3} \frac{\dot{Q}_{\rm mb}}{H},\\ 
&\frac{d T_{\rm C}}{d \log a} =-2 T_{\rm C}+\frac{2}{3} \frac{\dot{Q}_{\rm Cm}}{H},\\ 
&\frac{d V_{\rm { bm }}}{d \log a} =-V_{\rm { bm }}-\frac{D_{\rm bm}}{H},\\
&\frac{d V_{\rm { mC }}}{d \log a} =-V_{\rm { mC }}-\frac{D_{\rm mC}}{H} ,
\end{align}
where $\Gamma_{\rm Comp}$ is the Compton rate of interaction between photons and baryons, which is dominated by electrons. We have neglected the Compton heating term from $T_\text{m}$: this is justified if the thermal decoupling from the CMB occurs before recombination at $z \sim 1100$, so that after this redshift, the Compton heating rate is negligible compared to the adiabatic cooling rate.
The redshift of thermal decoupling for mDM occurs at~\cite{Liu:2018uzy}
\begin{alignat}{1}
    (1 + z)_\text{td} \approx \left(\frac{45 m_\text{m} H_0 \sqrt{\Omega_m}}{4 \pi^2 Q^4 \sigma_T T_{\gamma,0}^4}\right)^{2/5} \,,
\end{alignat}
where $T_{\gamma,0}$ is the CMB temperature today. We therefore see that for thermal decoupling to occur at $z \gtrsim 1100$, we require
\begin{alignat}{1}
    Q \lesssim 0.35 \left(\frac{m_\text{m}}{\SI{}{MeV}}\right)^{1/4} \,,
\end{alignat}
which is always satisfied when the contours shown in Fig.~\ref{fig:money_milli} lie within the current experimentally allowed region. 

Using the formalism already introduced in App.~\ref{sec:21cm} we can write the above equations more explicitly in terms of a few simple 1D integrals:
\begin{align} \label{eq:Tdot_full}
&\frac{d T_{\rm b}}{d \log a} +2 T_{\rm b}=\frac{2f_{\rm m}\rho_{\rm DM}}{3H(1 + x_e + \mathcal{F}_\text{He})} \sum_{ j}\frac{x_{j}\mu_{j\rm m}}{m_{\rm m}+m_{ j}}\left[I_{j\rm m}^D +\frac{T_{\rm m}-T_{\rm b}}{m_{\rm m}u_{j\mathrm{m}}^{2}}I_{j\rm m}^T\right]+\frac{\Gamma _{\rm Comp}}{H}\left(T_\gamma-T_{\rm b}\right),\\ 
&\frac{d T_{\rm C}}{d \log a} +2 T_{\rm C}=\frac{2f_{\rm m}\rho_{\rm DM}}{3H} \frac{\mu_{\rm mC}}{m_{\rm m}+m_{\rm C}}\left[I_{\rm mC}^D +\frac{T_{\rm m}-T_{\rm C}}{m_{\rm m}u_{\mathrm{mC}}^{2}}I_{\rm mC}^T\right] ,\\ 
&\frac{d T_{\rm m}}{d \log a} +2 T_{\rm m}=\frac{2(1-f_{\rm m})\rho_{\rm DM}}{3H} \frac{\mu_{\rm mC}}{m_{\rm m}+m_{\rm C}}\left[I_{\rm mC}^D +\frac{T_{\rm C}-T_{\rm m}}{m_{\rm C}u_{\mathrm{mC}}^{2}}I_{\rm mC}^T\right]+\frac{2}{3H}  \sum_{ j}  \frac{n_j m_j \mu_{j\rm m}}{m_{\rm m}+m_{ j}}\left[I_{j\rm m}^D +\frac{T_{\rm b}-T_{\rm m}}{m_j u_{j\mathrm{m}}^{2}}I_{j\rm m}^T\right], \label{eq:Tm_evol}\\ 
&\frac{d V_{\rm { bm }}}{d \log a} +V_{\rm { bm }}=-\left(\frac{\rho_{\rm m}}{\rho_{\rm b}}+1\right)\sum_{j}\frac{\rho_{j}}{m_{\rm m}+m_j}\frac{I^D_{j\rm m}}{HV_{\rm bm}}+\frac{\rho_{\rm C}}{m_{\rm m}+m_{\rm C}}\frac{I^D_{\rm mC}}{HV_{\rm mC}},\\
&\frac{d V_{\rm { mC }}}{d \log a} + V_{\rm { mC }}=-\frac{\rho_{\rm m}+\rho_{\rm C}}{m_{\rm m}+m_{\rm C}}\frac{I^D_{\rm mC}}{HV_{\rm mC}}+\sum_{j}\frac{\rho_{j}}{m_{\rm m}+m_j}\frac{I^D_{j\rm m}}{HV_{\rm bm}},\\
&\frac{d x_{e}}{d\log a}=-\frac{C}{H}\left(n_{H} \mathcal{A}_{B} x_{e}^{2}-4\left(1-x_{e}\right) \mathcal{B}_{B} e^{3 E_{0} /\left(4 T_{\gamma}\right)}\right).
\label{eq:Tdot_full_last}
\end{align}
The index $j$ runs over all the components of the baryonic bath that interact with the mDM, with $m_j$ and $\mu_{j\rm m}$ being the mass of the $j$th particle and the $j$-m reduced mass respectively. The quantities $I^{T/D}_{XX}$ are the thermal averaged integrals defined in Eqs.~(\ref{eq:int1}) and~(\ref{eq:int2}), using the appropriate form factors from App.~\ref{sec:study_case}; these quantities may also carry an index $j$ for their dependence on $u_j^2=T_{\rm b}/m_j+T_{\rm m}/m_{\rm m}$. The sixth equation accounts for the dynamics of the ionized fraction $x_e$ that is coupled to the other variables we solve for. The temperature of the CMB photons is assumed to be decoupled from matter for all practical purposes and redshifts as a relativistic fluid throughout.

It is instructive to study some limiting cases to get some intuition for the set of equations above. The mDM-CDM interaction is controlled by $\alpha_\text{C} \alpha_\text{m}$ while the mDM-baryon interaction is controlled by $Q^2\alpha_{\text{EM}}^2$. In the limit where $\alpha_\text{C} \alpha_\text{m}\to 0$, the equations above reduce to the two-fluid system presented in Appendix~\ref{sec:21cm} and the analysis is the usual one performed in recent literature~\cite{Munoz:2018pzp,Berlin:2018sjs,Barkana:2018cct,Kovetz:2018zan}. Conversely, in the limit $\alpha_\text{C} \alpha_\text{m}\to \infty$, mDM and CDM are tightly coupled and the equations reduce again to a two-fluid system. If we neglect the mass difference between $m_\text{C}$ and  $m_\text{m}$, the resulting equations are exactly those of 100\% millicharged DM with an effective DM electric charge equal to $\sqrt{f_{\rm m}}Q$.
The first limit correspond to the standard mDM explanation of the EDGES measurement which has been shown to be challenged by a number of cosmological, astrophysical and ground based constraints~\cite{Munoz:2018pzp,Berlin:2018sjs,Barkana:2018cct,Kovetz:2018zan,Creque-Sarbinowski:2019mcm}. The second one is ruled out by the CMB constraints alone \cite{Dvorkin:2013cea,Kovetz:2018zan,Boddy:2018wzy}.

\begin{figure*}
\includegraphics[width=0.56\textwidth]{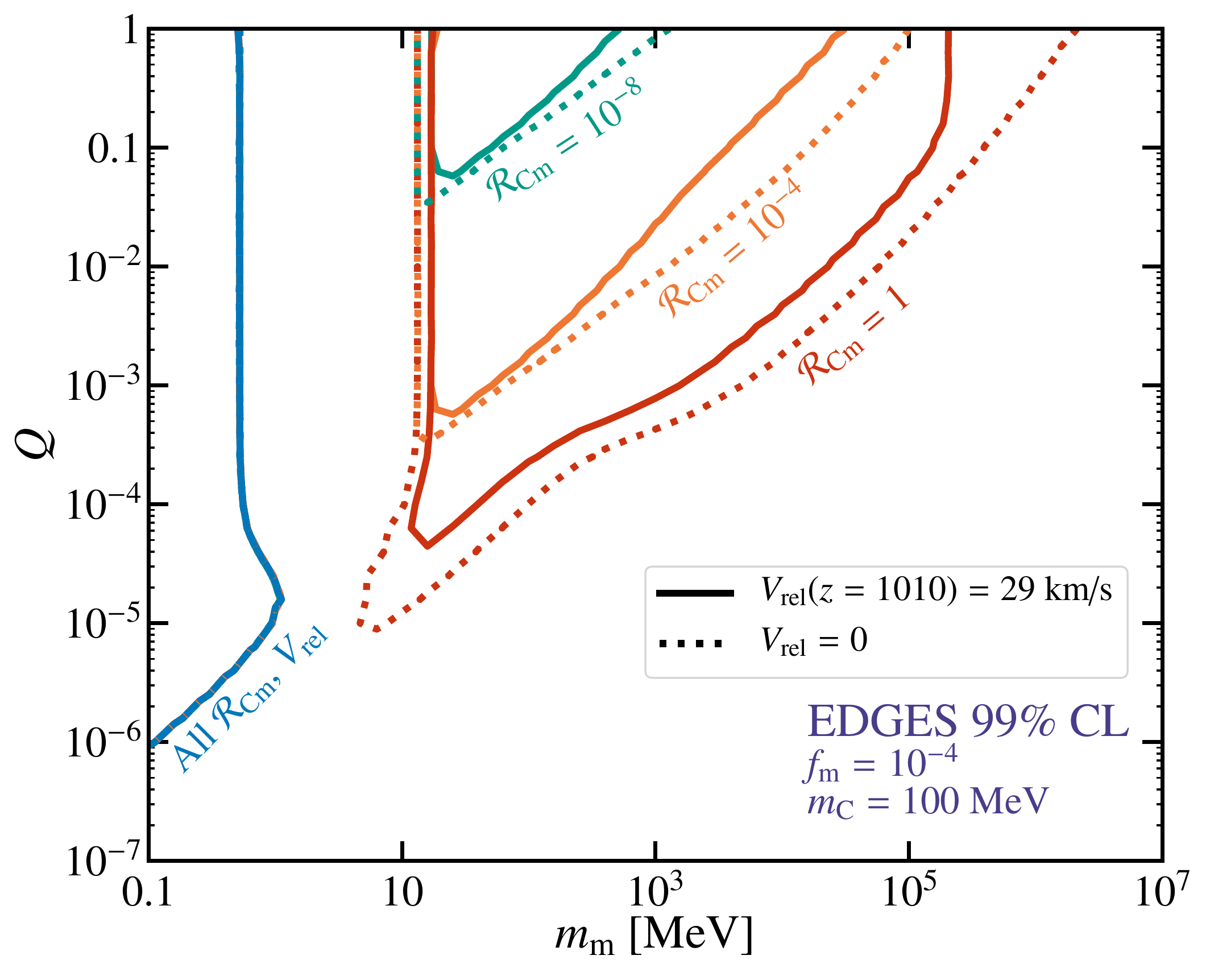}
\caption{Contours of $Q$ and $m_\text{m}$ that produce sufficient cooling of baryons to satisfy the EDGES 99\% CL limit for different values of $\mathcal{R}_\text{Cm}$, defined in Eq.~(\ref{eq:C_param}), while also remaining consistent with the CMB bounds in Eq.~(\ref{eq:cmbconstrain}). Here, $f_\text{m} = 10^{-4}$, $m_\text{C} = \SI{100}{\mega\eV}$. We have also set $V_\text{mb} = 0$ initially, and show contours for both $V_\text{mC} = \SI{29}{\kilo\meter\per\second}$ at $z = 1010$ (\textbf{solid}) and $V_\text{mC} = 0$ (\textbf{dotted}). For all values of $\mathcal{R}_\text{Cm}$ and initial conditions, the contour shown in the region where $m_\text{m} < \SI{1}{\mega\eV}$ (\textbf{blue}) is allowed, corresponding to the case where the mDM fluid alone can cool the baryons sufficiently (at such small $m_\text{m}$, $V_\text{mC}$ does not significantly affect the thermal history~\cite{Munoz:2015bca}). In the region $\SI{1}{\mega\eV} \lesssim m_\text{m} \lesssim \SI{10}{\mega\eV}$, the mDM fluid does not have sufficient heat capacity to cool baryons alone, but also does not recouple to the CDM fluid sufficiently early. As a consequence, the CDM fluid is unable to assist in cooling. Our setup opens up the parameter space for $m_\text{m} \gtrsim \SI{10}{\mega\eV}$, when the mDM fluid cools below the baryon temperature and equilibrates with the CDM fluid after recombination, allowing for an enhanced cooling of baryons. Contours representing parameter values for $\mathcal{R}_\text{Cm} = 10^{-8}$ (\textbf{green}), $10^{-4}$ (\textbf{orange}) and $1$ (\textbf{red}) are shown, showing that a large range of $\alpha_\text{C} \alpha_\text{m}$ is possible in our model.} 
\label{fig:qeff}
\end{figure*}

\begin{figure*}[t]
\centering
\includegraphics[width=\textwidth]{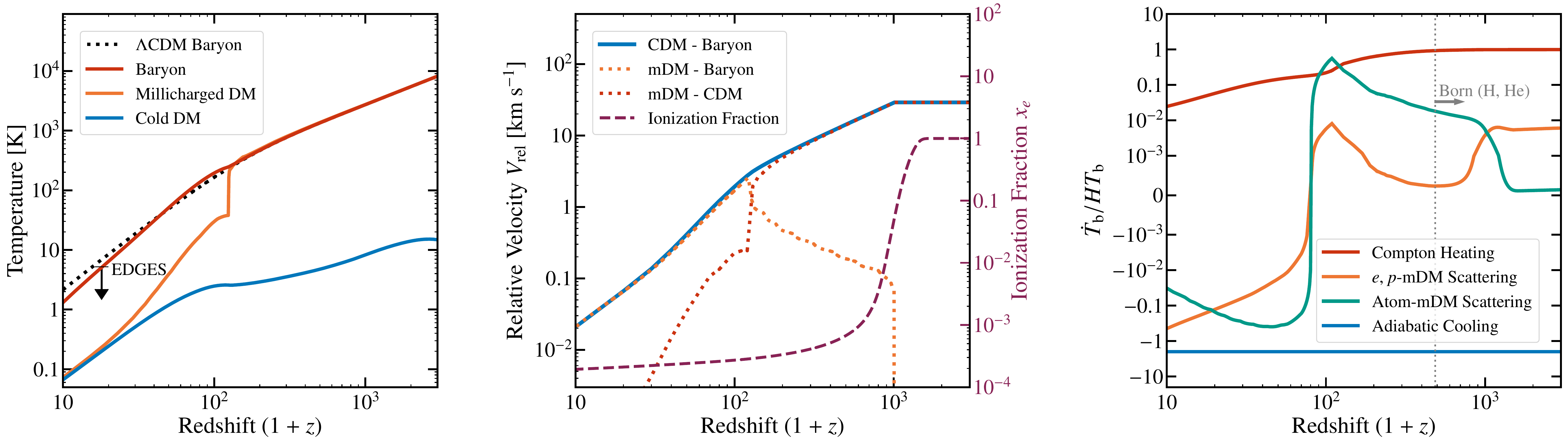}
\caption{Evolution of the fluid temperatures ({\bf left}), bulk relative velocities and ionization ({\bf center}), together with the heating/cooling rates $\dot{T}_\text{b}/ HT_\text{b}$ of relevant processes ({\bf right}) for benchmark parameters given by $Q=1.2 \times 10^{-3}$, $m_\text{m} = 2\text{ GeV}$ and $f_\text{m} = 10^{-4}$, with $m_\text{C}=100\text{ MeV}$ and $\alpha_\text{m} \alpha_\text{C}=2 \times 10^{-15}$ setting the interactions between mDM and CDM. Compared to Fig.~\ref{fig:temp_ev}, the bulk relative velocity between mDM and CDM ({\bf center, dotted red}) $V_\text{mC}$ results in an initial phase of heating as the mDM fluid experiences a drag from the CDM fluid. This drag force is sufficient for $V_\text{mC} \sim 0$ by $z \sim 200$ and for a bulk relative velocity between mDM and baryons ({\bf center, dotted orange}) $V_\text{bm}$ to develop. At this point, the mDM fluid begins to cool the baryons, satisfying the EDGES result for this choice of parameters. The bulk relative velocity between CDM and baryons $V_\text{bm} + V_\text{mC}$ ({\bf center, solid blue}) evolves more gradually, redshifting as $1+z$ just after recombination as in $\Lambda$CDM cosmology. It is fixed at a constant value of \SI{29}{\kilo\meter\per\second} prior to kinetic decoupling at $z = 1010$~\cite{Tseliakhovich:2010bj}.}
\label{fig:temp_ev_Vrel}
\end{figure*}

The intermediate scenario is the one of interest for this paper: 
\begin{itemize}
\item $\alpha_\text{C} \alpha_\text{m}$ must be small enough to not be excluded by the CMB constraints on the CDM momentum drag on the baryonic fluid. In App.~\ref{sub:constraints} we will show how the existing CMB bounds on mDM \cite{Dvorkin:2013cea,Kovetz:2018zan,Boddy:2018wzy} can be extended with a simple rescaling to our scenario under the assumptions that mDM is thermally coupled to the baryons before recombination, while CDM is decoupled. We will work under this assumption to simplify the CMB analysis of our paper, deferring a thorough study of the exact CMB limits for future studies. In our treatment, the maximal value of the couplings, $\alpha_\text{C} \alpha_\text{m}\vert_{\text{max}}$,  must also ensure that mDM remains tightly coupled to baryons, as discussed later in App.~\ref{sub:constraints}. 
\item $\alpha_\text{C} \alpha_\text{m}$ must be large enough such that CDM drives the cooling of the baryonic bath necessary to explain the EDGES observation. $\alpha_\text{C} \alpha_\text{m}$ is then fixed for a given value of $Q$, $m_{\text{m}}$ and $m_{\text{C}}$. For the same choice of couplings, in many regions of the parameter space,  mDM will couple back to the CDM bath after recombination (when  $x_e$ drops rapidly). 
\end{itemize}
These two conditions identify an interval for $\alpha_\text{C} \alpha_\text{m}$ whose boundaries can be parametrized by the ratio 
\begin{alignat}{1}
\mathcal{R}_{\text{C}\text{m}}\equiv\frac{\alpha_\text{C} \alpha_\text{m}}{\alpha_\text{C} \alpha_\text{m}\vert_{\text{max}}} \,.
    \label{eq:C_param}
\end{alignat}
$\mathcal{R}_{\text{C}\text{m}}=0$ corresponds to $\alpha_\text{C} \alpha_\text{m} \to 0$ which brings us back to the standard mDM scenario, while $\mathcal{R}_{\text{C}\text{m}}=1$ corresponds to the boundary of the parameter space, where $\alpha_\text{C} \alpha_\text{m}$ required to cool the baryons is equal to the maximum allowed value by CMB constraints in the tight coupling approximation between mDM and baryons. Fig.~\ref{fig:qeff} shows contours of $\mathcal{R}_{\text{C}\text{m}}$ on the $m_\text{m}$ -- $Q$ plane with $f_\text{m} = 10^{-4}$ and $m_\text{C}=100\text{ MeV}$. Notice that $\mathcal{R}_{\text{C}\text{m}}$ can also be used as a measure of the allowed model space of our construction, as we can see for $f_\text{m}=10^{-4}$ most of the parameter space allows for $\mathcal{R}_{\text{C}\text{m}}\ll 1$, which corresponds to a wide range of possible choices for $\alpha_\text{m} \alpha_\text{C} $ to fit the EDGES signal.

Neglecting the drag term between the mDM and the CDM fluid, for $\mathcal{R}_{\text{C}\text{m}}=0$ our setup reproduces the result with no CDM bath, which is consistent with the results in Ref.~\cite{Munoz:2018pzp,Berlin:2018sjs}. As $\mathcal{R}_{\text{C}\text{m}}$ approaches one, we approach the contour for fixed $f_\text{m}$ shown in Fig.~\ref{fig:money_milli}, demonstrating the gradual transition from the pure millicharged model with no CDM bath considered prior to this paper and this work. Including the drag force, the initial relative velocity between the CDM and the mDM bath should drop below the thermal relative velocity at sufficiently high redshift, otherwise the heating induced by the drag force counteracts the thermal cooling making it impossible for mDM to couple back to CDM. This means a larger value of $Q$ is required at each value of $m_\text{m}$ to get sufficient cooling to explain the EDGES result. An example of a successful thermal history which accounts for the drag force is showed in Fig.~\ref{fig:temp_ev_Vrel}. The initial bulk relative velocity $V_{\text{mC}}$ between mDM and CDM drops rapidly around $z\simeq 200$ allowing the mDM to decouple from the baryonic bath and begin cooling baryons. From Fig.~\ref{fig:qeff}, we see that at fixed CDM mass, the tension between a large deceleration rate and the CMB constraint on $\alpha_{\text{m}}\alpha_{\text{C}}$ selects an interval of mDM masses. In the gap between the standard mDM solution and the region of parameter space opened up by our setup, the non-zero value of $\alpha_\text{C} \alpha_\text{m}$ required to cool the baryons sufficiently for the EDGES observation violates the upper bound imposed by the CMB constraints discussed below.

\subsection{Parameter Constraints}
\label{sub:constraints}

We discuss here the cosmological constraints on our setup. We first discuss the CMB constraints and how the existing analysis of the standard mDM case (without interactions with CDM) can be extended to our framework. We then comment on how the effects of annihilations can be neglected in our analysis. 

\subsubsection{CMB Constraints} 
The CMB power spectrum measurements can be used to constrain the interaction strength between the dark and the baryonic fluid~\cite{Dubovsky:2003yn,Dolgov:2013una,Dvorkin:2013cea,dePutter:2018xte,Kovetz:2018zan,Boddy:2018wzy}. In what follows we recast these CMB constraints to the particular setup considered in this work. 

The CMB constraint derived in  Refs.~\cite{Dubovsky:2003yn,Dolgov:2013una,Kovetz:2018zan} applies to the standard mDM case, where an mDM fraction $f_{\text{m}}$ of total DM density is assumed to be completely decoupled from the CDM component. In this case the upper bound on the mDM cross section with the baryons is roughly $\sigma_{\rm max}\lesssim \SI{1.7e-41}{\centi\meter\squared}$ as long as $f_{\text{m}}\gtrsim 4\times 10^{-3}$ \cite{Dvorkin:2013cea,Boddy:2018wzy,Kovetz:2018zan,dePutter:2018xte}. 

In the presences of mDM-CDM interactions we consider two opposite limiting cases leaving the treatment of the intermediate case for a more detailed analysis. 
\begin{itemize}  
\item \emph{mDM tightly coupled to CDM.} If the mDM fraction $f_{\rm m}$ of total DM density is tightly coupled to the CDM component before recombination the baryonic bath feels the drag force from the full DM bath but all the rates are multiplied by $f_{\rm m}$. This means that if the upper bound on the interaction cross section of pure mDM with the baryons was $\sigma_{\rm max}$, when mDM is tightly coupled to CDM the upper bound gets rescaled as $\sigma_{\rm max}/f_{\rm m}$. 
\item \emph{mDM tightly coupled to the baryons.} When the mDM fraction is tightly coupled to the baryonic fluid before recombination it can effectively be considered as a component of the baryonic fluid at CMB. The CDM interaction with the baryons is then dominated by the CDM interactions with the mDM component. Since the CMB constrains mainly the drag exerted on the baryonic fluid, we want to compute the drag of CDM on the mDM component. For $m_{\rm m}\gtrsim m_p$ the relative fluid velocity is dominated by the relative bulk motion so that following the derivation of Eq.~\eqref{eq:drag} and~\eqref{eq:V_to_0} one finds
\begin{equation}
    D_{\rm mC}=\left(1+\frac{\rho_{\mathrm{C}}}{\rho_{\mathrm{b}}+\rho_{\mathrm{m}}}\right) \frac{\rho_{\mathrm{m}}}{m_{\mathrm{m}}+m_{\mathrm{C}}} \sigma_{T}^{\mathrm{mC}} V_\text{mC}^2,
\end{equation}
where $\sigma_{T,\rm mC}$ is the transfer cross section for the CDM-mDM interaction. This result allows us to recast the constraint on the cross section from Ref.~\cite{Boddy:2018wzy} as 
\begin{equation}
    \sigma_{T}^{\mathrm{mC}}\left(V_\text{mC}\right) V_\text{mC}^4 \lesssim \SI{1.7e-41}{\centi\meter\squared} \left( \frac{m_{\mathrm{C}}+m_{\mathrm{m}}}{m_{p}} \right) \left(1+\frac{\Omega_{\mathrm{b}}}{f_{\mathrm{m}}\Omega_{\mathrm{DM}}}\right) .
\end{equation}
Notice that forecast constraints from CMB stage-4, according to Ref.~\cite{Boddy:2018wzy}, would improve this constraint only by a factor of 2. In terms of coupling constants, this reads
\begin{alignat}{1}
    \alpha_\text{C} \alpha_\text{m} \lesssim 7.4 \times 10^{-19} \left(\frac{m_\text{m}}{\SI{1}{\giga\eV}}\right)^2 \left(\frac{m_\text{C}}{\SI{100}{\mega\eV}}\right)^2 \left(\frac{\SI{1}{\giga\eV}}{m_\text{C} + m_\text{m}}\right) \left(1 + \frac{\Omega_\text{b}}{f_\text{m} \Omega_\text{DM}}\right) \,.
    \label{eq:CMB_coupling_limit}
\end{alignat}
\end{itemize}

In order for mDM to be tightly coupled to baryons before recombination as assumed in deriving Eq.~\eqref{eq:CMB_coupling_limit}, we need the energy transfer rate between baryons and mDM to be significantly larger than the energy transfer rate between CDM and mDM. This gives an upper bound on $\alpha_\text{C} \alpha_\text{m}$ for a given value of $Q$. In order to get an analytical approximation for this, we can observe from Eq.~(\ref{eq:Tm_evol}) that $T_m \sim T_b$ if the heat transfer terms from the CDM fluid is much smaller than the heat transfer terms from the baryons with $T_m \to 0$; explicitly, this means we require
\begin{alignat}{1}
    \frac{\rho_\text{DM} \mu_\text{mC}}{m_\text{m} + m_\text{C}} I_\text{mC}^D \ll \frac{\rho_\text{b} \mu_\text{mp}}{m_\text{m} + m_\text{p}} \frac{T_\text{b}}{m_\text{p} u_\text{mp}^2} I_\text{mp}^T \,,
\end{alignat}
where we have assumed that the energy transfer from CDM to mDM is dominated by the bulk relative velocity for simplicity. Using the limiting expressions found in Eqs.~(\ref{eq:V_to_0})--(\ref{eq:V_to_inf}) and neglecting log factors, we find
\begin{alignat}{1}
    \alpha_\text{C} \alpha_\text{m} \ll 6.3 \times 10^{-6} Q^2 \left(\frac{m_\text{C}}{\SI{100}{\mega\eV}}\right)
    \label{eq:tight_coupling_requirement}
\end{alignat}
For concreteness, we take the maximum value of $\alpha_\text{C} \alpha_\text{m}$ for tight coupling between mDM and baryons to be
\begin{alignat}{1}
    \alpha_\text{C} \alpha_\text{m}\vert_\text{max} \ll 6.3 \times 10^{-7} Q^2 \left(\frac{m_\text{C}}{\SI{100}{\mega\eV}}\right) \,.
\end{alignat}
This has been found numerically to be sufficient to ensure that $T_\text{m} \sim T_\text{b}$ at recombination. In reality, this is simply a heuristic bound to guarantee the accuracy of recasting of the CMB limits; a full analysis would likely relax our assumptions significantly.

\subsubsection{Dark Matter Annihilation}

Annihilation of either mDM or CDM particles can lead to energy injection into any of the three fluids we consider here. In what follows we argue that the effects of the various annihilation processes are always small in the parameter space of interest. 

\begin{itemize}

    \item \textit{mDM/CDM to SM Annihilations}. These processes produce high-energy particles that ultimately heat up the baryons, possibly counteracting any cooling processes~\cite{Liu:2018uzy}. To determine if this has a significant effect on the baryon temperature evolution, we can simply compare the energy injection rate with the cooling due to adiabatic expansion.

	For mDM annihilations into SM particles, we can take the annihilation cross section to be $\langle \sigma v \rangle \sim Q^2 \alpha_\text{EM}^2 / m_\text{m}^2$. The energy injected per baryon per unit time is given by
	\begin{alignat}{1}
	    \frac{f_\text{m}^2 \rho_\text{DM}^2}{n_\text{H} (1 + x_e + \mathcal{F}_\text{He})} \frac{\langle \sigma v \rangle}{m_\text{m}} \sim  \SI{e-34}{\giga \eV \per \second} \left( \frac{f_\text{m}}{10^{-4}}\right)^2 \left(\frac{\SI{1}{\giga \eV}}{m_\text{m}}\right)^3 (1+z)^3 Q^2\,.
	\end{alignat}
	This should be compared to the adiabatic expansion cooling rate of the baryons, 
	\begin{alignat}{1}
	    H T_b \sim \SI{e-27}{\giga\eV \per \second} \left(\frac{T_\text{b}}{\SI{1}{\eV}}\right) (1 + z)^{3/2} \,.
	\end{alignat}
	We can therefore neglect mDM annihilations into SM particles if
	\begin{alignat}{1}
	    Q \ll 20 \left(\frac{10^{-4}}{f_\text{m}}\right) \left(\frac{m_\text{m}}{\SI{1}{\giga\eV}}\right)^{3/2} \left(\frac{T_\text{b}}{\SI{1}{\eV}}\right)^{1/2} \left(\frac{10^3}{1+z}\right)^{3/4} \,.
	    \label{eq:Q_to_neglect_mDM_to_SM}
	\end{alignat}
	Prior to recombination, the baryon fluid is coupled tightly to photons, and any energy injection at this time will have no effect on the baryon temperature. It is easy to check that Eq.~(\ref{eq:Q_to_neglect_mDM_to_SM}) is satisfied after recombination across the entire allowed parameter space in Fig.~\ref{fig:money_milli}.

    For CDM, the annihilation rate is suppressed by loop factors and the dark sector couplings $\alpha_\text{C} \alpha_\text{m}$, but the energy density is much larger. Taking $\langle \sigma v \rangle \sim Q^2 \alpha_\text{EM}^2 \alpha_\text{C} \alpha_\text{m} / m_\text{C}^2$ and substituting the CMB upper limit on $\alpha_\text{C} \alpha_\text{m}$ in Eq.~\eqref{eq:CMB_coupling_limit} gives
    \begin{alignat}{1}
        Q \ll 300 \left(\frac{f_\text{m}}{10^{-4}}\right)^{1/2} \left(\frac{\SI{1}{\giga\eV}}{m_\text{m}}\right) \left(\frac{m_\text{C}}{\SI{100}{\mega\eV}}\right)^{1/2} \left(\frac{m_\text{C} + m_\text{m}}{\SI{1}{\giga\eV}}\right)^{1/2} \left(\frac{T_\text{b}}{\SI{1}{\eV}}\right)^{1/2} \left(\frac{10^3}{1+z}\right)^{3/4} \,,
    \label{eq:Q_to_neglect_CDM_to_SM}
    \end{alignat}
    which is again satisfied in the parameter space of interest. The estimates shown in Eqs.~(\ref{eq:Q_to_neglect_mDM_to_SM}) and~(\ref{eq:Q_to_neglect_CDM_to_SM}) are highly conservative, assuming instantaneous deposition of all energy produced from all annihilations. We therefore find it reasonable to neglect all annihilations to SM particles throughout.

    \item \textit{mDM (CDM) to CDM (mDM) annihilations}.
   
    A pair of mDM particles can annihilate into a pair of CDM particles or vice versa, depending on the masses of the respective particles; this may inject energy into either fluid. Suppose $m_\text{C} > m_\text{m}$, so that CDM particles annihilate into mDM particles with a cross section $\langle \sigma v \rangle_\text{C} \sim \alpha_\text{C} \alpha_\text{m} / m_\text{C}^2$. As we argued above, these annihilations would significantly influence the temperature evolution of the CDM bath if the energy injection rate were comparable to the energy lost due to Hubble expansion (the potential effects on the mDM bath is much smaller since the mDM bath has a higher temperature, leading to a significantly larger adiabatic cooling rate). The maximum amount of energy injected from each annihilation event is $m_\text{C}$. Therefore, the energy injection from CDM annihilations into mDM particles is much smaller than the Hubble expansion cooling if $n_\text{C} \langle \sigma v \rangle_\text{C} m_\text{C} \ll H T_\text{C}$. Adopting the maximum value of $\alpha_\text{C} \alpha_\text{m}$ from the CMB limits derived below in Eq.~(\ref{eq:CMB_coupling_limit}), this gives
    \begin{alignat}{1}
        T_\text{C} (z) \gg \SI{5e-7}{\eV}  \left(\frac{m_\text{m}}{\SI{10}{\mega\eV}}\right)^2 \left(\frac{\SI{10}{\mega\eV}}{m_\text{C} + m_\text{m}}\right) \left( \frac{10^{-4}}{f_\text{m}}\right) \left( \frac{1 + z}{10^3} \right)^{3/2} \,.
        \label{eq:T_C_annihilation_limit}
    \end{alignat}
    This is easily satisfied for $f_\text{m} = 10^{-4}$ and for the parameter choices in Fig.~\ref{fig:money_milli}, with $T_\text{C}(z_\text{rec})$ typically reaching \SI{e-3}{\eV} when solving Eqs.~(\ref{eq:Tdot_full}) --~(\ref{eq:Tdot_full_last}) numerically.  Eq.~(\ref{eq:T_C_annihilation_limit}) is an extremely conservative requirement, since we have neglected the efficiency with which high-energy mDM particles deposit energy into the CDM fluid, which is highly model dependent, but may reduce the energy deposited significantly.  We stress that this statement is independent of the initial conditions for any set of parameters that can explain the EDGES signal: the coupling between mDM and CDM is strong enough such that the value of $T_\text{C}$ evolves rapidly to the same level around recombination, regardless of the initial conditions specified.

    If on the other hand $m_\text{m} > m_\text{C}$, then the energy injection rate from mDM annihilations is further suppressed by $f_\text{m}^2$, and the same analysis leads to even weaker requirements on $T_\text{C}(z)$. 

\end{itemize}

\section{A Vector Portal between Millicharged and Cold Dark Matter}
\label{sec:bathmodels}
\begin{figure*}[t!]
\centering
\includegraphics[width=0.5\textwidth]{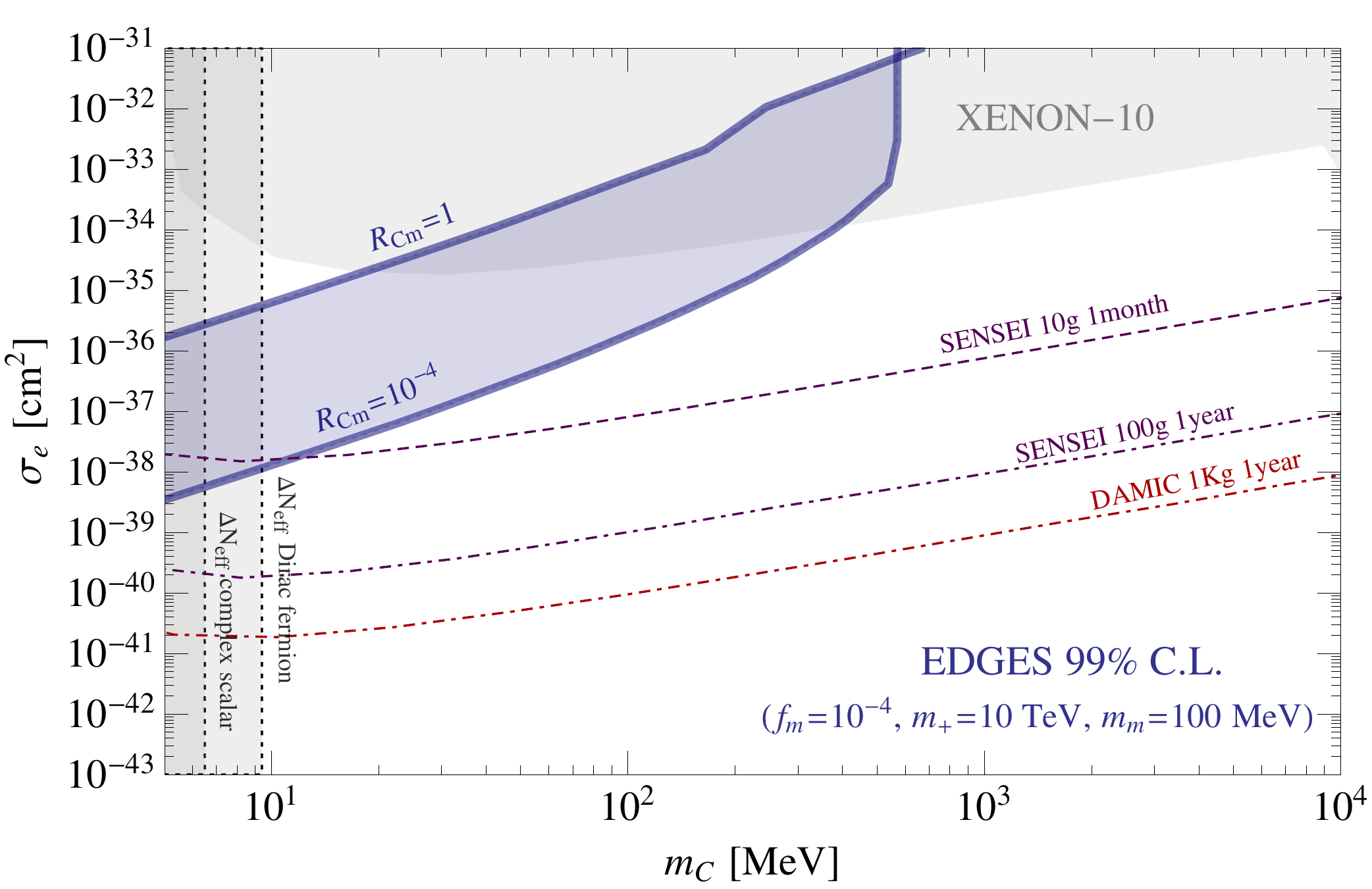}
\caption{The thick blue band encloses the predictions for the direct detection cross section of the CDM component fitting EDGES for fixed mDM mass $m_\text{m}=100\text{ MeV}$ and mDM fraction $f_\text{m}=10^{-4}$ lowering $\alpha_\text{m} \alpha_\text{C}$ from its maximal value satisfying the CMB constraint down to the entire interval allowed by fitting EDGES (see Fig.~\ref{fig:qeff}). The cross section does not change by more than $2$ orders of magnitude by lowering the coupling up to eight orders of magnitude, showing the robustness of our direct detection prediction. Details about the present constraints and prospects can be found in the caption of Fig.~\ref{fig:money_CDM_vector}. }
\label{fig:direct_alpha}
\end{figure*}

We illustrate here two simple models realizing the mDM-CDM setup of Fig.~\ref{fig:cartoon}, namely the vector portal. In this simple case we determine the cross section of CDM with the SM, which is induced by loops of mDM particles. The prospects for direct detection are discussed in Sec.~\ref{sec:pheno}.  

We first consider a theory with two vector currents: one associated with mDM coupled to the SM photon and the other associated with the new dark vector boson $V$ coupling mDM to CDM. The Lagrangian of the model is (Lorentz indices are suppressed) 
\begin{equation}
\mathcal{L}= \frac{1}{4}F^2+ \frac{1}{4}F^2_V+ \frac{\kappa}{4}F F_V+ e A J_{\text{EM}}+g_V V J_V + {\cal L}_m\,,
\end{equation}
where $\mathcal{L}_m$ encodes vector-like masses for the matter content which we'll take to be two Dirac states (written in Weyl notations): the mDM, $\chi$-$\tilde{\chi}$, and the CDM, $\psi$-$\tilde{\psi}$.
The generated mixing operator between the two field strengths depends on the structure of the matter currents, which in turn affect the structure of the gauge invariant masses for the matter fields. We assume the matter fields to be fermions for concreteness, but we do not expect our arguments to be modified in the scalar case. In the simplest model, we can write 
\begin{align}
& J_{\text{EM}}= Q(\chi\sigma\chi^\dagger-\tilde{\chi}\sigma\tilde{\chi}^\dagger)+J_{\text{EM}}^{\text{SM}}\ ,\label{eq:basicem}\\
& J_{V}= q_\text{m}(\chi\sigma\chi^\dagger-\tilde{\chi}\sigma\tilde{\chi}^\dagger)+q_\text{C}(\psi\sigma\psi^\dagger-\tilde{\psi}\sigma\tilde{\psi}^\dagger)\ ,\\
&\mathcal{L}_m=m_\text{m} \tilde{\chi}\chi+m_\text{C}\tilde{\psi}\psi\,,\label{eq:basicmasses}
\end{align}
 Notice that in this simple case there is a residual symmetry acting on the fermions and the gauge fields under which the full Lagrangian is invariant: 
\begin{align}
\chi\to \tilde{\chi}\quad ,\ \psi\to \tilde{\psi}\quad ,\ A\to -A \quad ,\ V\to -V \ .
\end{align} 
Computing a loop of $\chi$'s we get the natural size of the mixing parameter $\kappa$ 
\begin{align}
\kappa(q^2)&=\frac{Q q_\text{m} e g_V}{4\pi^2}\int_0^1 dx \, 3x(x-1) \log\left[\frac{\Lambda^2}{m_\text{m} ^2+q^2(x-1)x}\right]\simeq \frac{Q q_\text{m} e g_V}{4\pi^2}\log \frac{\Lambda}{m_\text{m} } \,,
\end{align}
where we expanded at low momenta with respect to the mDM masses and the mixing is logarithmically sensitive to the cut-off $\Lambda$. In this theory, the direct detection cross section of CDM with electrons reads (up to $\mathcal{O}(1)$ logarithmic terms)
\begin{equation}
\sigma_{\text{C}}^{e}\simeq\frac{2\pi\kappa^2\alpha_\text{C} \alpha_{\text{EM}}}{\mu^2_e v^4}\simeq \frac{2Q^2\alpha_\text{C} \alpha_\text{m} \alpha_{\text{EM}}^2}{\pi\mu^2_e v^4} \,,\label{eq:vectorstd}
\end{equation}   
where we defined 
\begin{equation}
\alpha_\text{m} \equiv \frac{g_\text{m}^2}{4\pi}=\frac{q_\text{m}^2g_V^2}{4\pi}\qquad ,\quad \alpha_\text{C} \equiv \frac{g_\text{C}^2}{4\pi}= \frac{q_\text{C}^2g_V^2}{4\pi} \,.
\end{equation}
For the region of couplings required to fit the EDGES result, this cross section is within the reach of present/near-future direct detection experiments as we show in the left panel Fig.~\ref{fig:money_CDM_vector}  in Sec.~\ref{sec:pheno}.  

Considering the next-to-minimal model we can see that the mixing between CDM and the SM can be arbitrarily suppressed via an accidental symmetry acting on the fermions and the gauge fields. The model introduces two Dirac-like pairs $\chi_{\pm}$-$\tilde{\chi}_{\pm}$ in the mDM sector which have opposite charges $\pm q_\text{m}$ under the dark gauge field $V$. The vector currents and the gauge invariant masses are in this case:
\begin{align}
& J_{\text{EM}}= Q(\chi_+\sigma\chi^\dagger_+-\tilde{\chi}_+\sigma\tilde{\chi}^\dagger_+)+Q(\chi_-\sigma\chi^\dagger_--\tilde{\chi}_-\sigma\tilde{\chi}^\dagger_-)\ ,\\
& J_{V}= q_\text{m}(\chi_+\sigma\chi^\dagger_+-\tilde{\chi}_+\sigma\tilde{\chi}^\dagger_+)-q_\text{m}(\chi_-\sigma\chi^\dagger_--\tilde{\chi}_-\sigma\tilde{\chi}^\dagger_-)+q_\text{C}(\psi\sigma\psi^\dagger-\tilde{\psi}\sigma\tilde{\psi}^\dagger)\ ,\\
&\mathcal{L}_m=m_{+}\tilde{\chi}_+\chi_++m_{-}\tilde{\chi}_-\chi_-+m_\text{C}\tilde{\psi}\psi\ .
\end{align}
In this theory there is now an accidental symmetry under which only the mixing operator $F F_V$ is odd:
\begin{align}
\chi_+\to \chi_-\quad ,\ \tilde{\chi}_+\to \tilde{\chi}_-\quad ,\ \psi\to \tilde{\psi}\quad ,\ A\to A \quad ,\ V\to -V\ .
\end{align} 
This symmetry is only softly broken by the mass difference between the two mDM Dirac-like pairs
\begin{equation}
\Delta m_\text{m} \equiv m_{-}-m_{+}\ .
\end{equation} 
For $\Delta m_\text{m}=0$, this symmetry is exact so we should find $\kappa=0$. By explicit computation we find indeed 
\begin{align}
\kappa(q^2)&=\frac{Q q_\text{m} e g_V}{4\pi^2}\int_0^1 dx  \, 3x(x-1) \log\left[\frac{m_{-}^2+q^2(x-1)x}{m_{+}^2+q^2(x-1)x}\right]\simeq \frac{Q q_\text{m} e g_V}{4\pi^2}\log \frac{m_{-}}{m_{+}}\simeq  \frac{Q q_\text{m} e g_V}{4\pi^2}\frac{\Delta m_\text{m} }{m_\text{m} }\ .
\end{align}
The corresponding cross section reads  
\begin{equation}
\sigma_{\text{C}}^{e}\simeq\frac{2Q^2\alpha_\text{C} \alpha_\text{m} \alpha_{\text{EM}}^2}{\pi\mu^2_e v^4}\left(\log \frac{m_{-}}{m_{+}}\right)^2\simeq \frac{2Q^2\alpha_\text{C} \alpha_\text{m} \alpha_{\text{EM}}^2}{\pi\mu^2_e v^4}\left(\frac{\Delta m_\text{m} }{m_\text{m} }\right)^2\ .
\end{equation}
We can therefore arbitrarily suppress the direct detection cross section of CDM by requiring a small splitting in between the dirac-like pairs of mDM. In the case of $m_{+}\gg m_{-}$ we recover the result in Eq.~\eqref{eq:vectorstd} if we identify the cut-off $\Lambda$ with $m_{+}$.

In the left/right panel of Fig.~\ref{fig:money_CDM_vector}  in Sec.~\ref{sec:pheno}, we show the prospects for discovering the CDM bath that explains EDGES for $m_{+}=10\text{ TeV}$/($\Delta m_\text{m}/m)=10^{-2}$. The direct detection cross section is fixed there by choosing $\alpha_\text{m}\alpha_{\text{C}}$ equal to its maximal value allowed by CMB. In Fig.~\ref{fig:direct_alpha} we show for a fixed choice of $m_{\text{m}}=100\text{ MeV}$ how much the direct detection cross section changes by lowering the value of $\alpha_\text{m}\alpha_{\text{C}}$. Since a smaller $\alpha_\text{m}\alpha_{\text{C}}$  should always be compensated by a higher $Q$, the direct detection cross section does not change by more than two orders of magnitude as shown in Fig.~\ref{fig:direct_alpha}.

\medskip

\end{document}